\documentclass[sigconf]{acmart}

\AtBeginDocument{%
  \providecommand\BibTeX{{%
    \normalfont B\kern-0.5em{\scshape i\kern-0.25em b}\kern-0.8em\TeX}}}

\copyrightyear{2024}
\acmYear{2024}
\setcopyright{acmlicensed}\acmConference[CHI '24]{Proceedings of the CHI Conference on Human Factors in Computing Systems}{May 11--16, 2024}{Honolulu, HI, USA}
\acmBooktitle{Proceedings of the CHI Conference on Human Factors in Computing Systems (CHI '24), May 11--16, 2024, Honolulu, HI, USA}
\acmDOI{10.1145/3613904.3642049}
\acmISBN{979-8-4007-0330-0/24/05}

\usepackage{color}
\usepackage{siunitx}
\usepackage{enumitem}
\usepackage{makecell}
\usepackage{array}
\usepackage{multirow}
\usepackage{ragged2e}
\usepackage{booktabs}
\usepackage{colortbl}
\usepackage{graphicx}
\usepackage{tabularray}
\usepackage{hyperref}
\usepackage{multicol}

\newcommand{\zq}[1]{{\color{black} #1}}

\begin{document}


\title[Designing User Acceptable Interaction for Situated Visualization in Public Environments]{Make Interaction Situated: Designing User Acceptable Interaction for Situated Visualization in Public Environments}  


\author{Qian Zhu}
\orcid{0000-0001-5108-3414}
\affiliation{
  \institution{The Hong Kong University of Science and Technology}
  \city{Hong Kong}
  \country{China}
}
\email{qian.zhu@connect.ust.hk}

\author{Zhuo Wang}
\orcid{0000-0003-4094-2708}
\affiliation{
  \institution{Xi’an Jiaotong Liverpool University}
  \city{Suzhou}
  \country{China}
}
\email{zhuohci01@gmail.com}

\author{Wei Zeng}
\authornote{Wei Zeng is the corresponding author.}
\orcid{0000-0002-5600-8824}
\affiliation{
  \institution{The Hong Kong University of Science and Technology (Guangzhou)}
  \city{Guangzhou}
  \country{China}
}
\affiliation{%
  \institution{The Hong Kong University of Science and Technology}
  \city{Hong Kong}
  \country{China}
}
\email{weizeng@hkust-gz.edu.cn}

\author{Wai Tong}
\orcid{0000-0001-9235-6095}
\affiliation{
  \institution{Texas A\&M University}
  \country{Texas, USA}
}
\email{wtong@tamu.edu}

\author{Weiyue Lin}
\orcid{0009-0000-7068-5752}
\affiliation{
  \institution{Peking University}
  \city{Beijing}
  \country{China}
}
\email{linweiyue@stu.pku.edu.cn}

\author{Xiaojuan Ma}
\orcid{0000-0002-9847-7784}
\affiliation{
  \institution{The Hong Kong University of Science and Technology}
  \city{Hong Kong}
  \country{China}
}
\email{mxj@cse.ust.hk}



\renewcommand{\shortauthors}{Qian Zhu, et al.}

\begin{abstract}
Situated visualization blends data into the real world to fulfill individuals' contextual information needs. 
However, interacting with situated visualization in public environments faces challenges posed by users' acceptance and contextual constraints. 
To explore appropriate interaction design, we first conduct a formative study to identify users' needs for data and interaction. Informed by the findings, we summarize appropriate interaction modalities with eye-based, hand-based and spatially-aware object interaction for situated visualization in public environments. 
Then, through an iterative design process with six users, we explore and implement interactive techniques for activating and analyzing with situated visualization. 
To assess the effectiveness and acceptance of these interactions, we integrate them into an AR prototype and conduct a within-subjects study in public scenarios using conventional hand-only interactions as the baseline.
The results show that participants preferred our prototype over the baseline, attributing their preference to the interactions being more acceptable, flexible, and practical in public.
\end{abstract}



\maketitle
\section{Introduction}
Recent technological advances in augmented reality (AR) have brought data visualization beyond the confines of desktops~\cite{lee2012beyond}, integrating it into the physical environment to facilitate in-situ data understanding and decision-making, such as for industrial manufacturing or daily shopping~\cite{becher2022situated, elsayed2016blended, xu2022arshopping}.
This results in a promising paradigm of data visualization -- situated visualization and analytics -- that integrates data with its physical referent or context for the analysis process~\cite{thomas2018situated, elsayed2016situated, white2009interaction, lee2023design, shin2023reality}, changing the way we perceive and interact with data in a real-world context.  

\begin{figure*}[htbp]
  \centering
  \includegraphics[width=\linewidth]{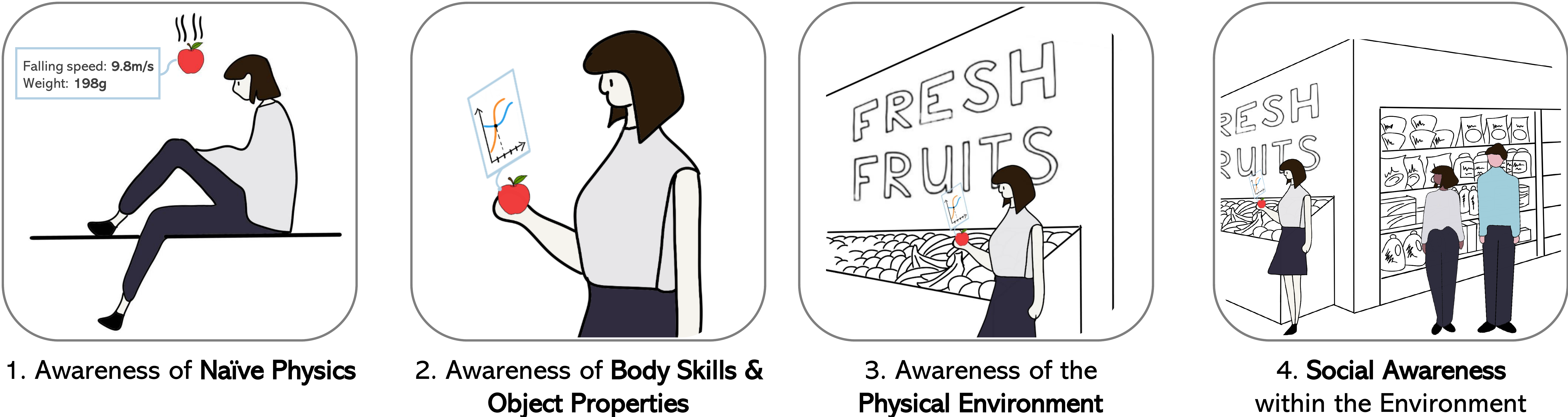}
  \caption{
  The four interaction themes for situated visualization from users' perspective, derived from the unified Reality-Based Interaction (RBI) framework~\cite{jacob2008reality}. 
  As users' awareness and understanding of the surrounding context gradually broaden, their adoption of the interactions with situated visualization can be more aligned with the real-world environment.}
  \label{fig:introMotivation}
\end{figure*}
Situated visualization enables a plausible future where data are seamlessly incorporated into the actual world to assist people in their daily decision-making and tasks~\cite{bach2017drawing,  bressa2022data}. 
In this process, interaction plays a critical role as it serves as the bridge connecting data, physical context, and end-users~\cite{thomas2018situated, white2009interaction}.
Existing research designed the interaction by leveraging the physical referent of the situated visualization~\cite{elsayed2016blended, satriadi2023active, satriadi2019augmented, white2009interaction}. 
For example, the earlier work by ElSayed et al.~\cite{elsayed2016blended} fused situated visualization with its physical referent, enabling users to interact through tangible touch inputs. Recent work by Satriadi et al.~\cite{satriadi2023active} employed the relationship between the situated visualization and its physical proxy to devise an interaction method for controlling the situated visualization through physical proxy manipulation.
However, prior research mainly focused on \zq{the physical aspects of the objects or environments to design interaction with situated visualizations (1, 2 and 3 in Fig.~\ref{fig:introMotivation}), neglecting users' awareness of others' presence within the environment (4 in Fig.~\ref{fig:introMotivation}).
People's acceptance and adoption of interaction with situated visualization can be influenced by their perception of both the environmental and social context~\cite{koelle2018acceptable, koelle2020social}.
Therefore, in this paper, we advocate for a broader consideration of context from the users' perspective compared to previous works when designing interactions with situated visualization, especially in public settings.
}

In a public context, we argue that interaction should align with the surrounding environment to ensure user acceptance and willingness to engage, as the interaction does not only occur between users and physical entities but also within the entire public environment~\cite{rico2010usable}. 
For instance, within a supermarket, users may be reluctant to continuously touch or shake a product to access information (e.g., shake menus in tangible AR~\cite{white2009shakemenu}), as this may damage the product, be perceived as socially inappropriate, or disturb the shopping experience of others~\cite{reeves2005designing}.
Therefore, designers need to consider not only the feasibility of interactions but also their user-perceived appropriateness for public use~\cite{hsieh2016designing, koelle2020social}). 
\zq{User acceptance, as a crucial perspective of social acceptability,} has been identified as an important aspect of interactive technology adoption in people's daily lives, as users may assess whether their motivation to engage with the interaction outweighs the potential risks of looking strange or committing a social blunder in public~\cite{rico2010usable, ahlstrom2014you, montero2010would}.
In addition, considering the environmental factors in advance could prevent investing substantial efforts in developing techniques that may be technically possible but could ultimately be rejected by users in real-world scenarios~\cite{montero2010would}. 

In this work, we aim to explore and design the appropriate interaction for situated visualization in public environments.
We target ordinary users who may exploit situated visualization to assist their daily tasks in public spaces and propose the following research questions:
\begin{itemize}
    \item \textbf{RQ1:} What are the requirements for interacting with situated visualization in public for daily tasks?
    \item \textbf{RQ2:} What design factors should be considered when designing interactions for situated visualization in public?
    \item \textbf{RQ3:} How effective and acceptable are the proposed interactions with situated visualization in public?
\end{itemize}

To answer \textbf{RQ1}, we conducted a formative study involving 12 participants in two common public scenarios – a library and a grocery store. 
We employed the contextual inquiry method to closely observe their natural practices, followed by in-depth interviews and in-situ demonstrations to uncover and elicit their needs~\cite{hartson2012ux}.
Considering the expansive design possibilities in situated visualization, we \zq{formulated the requirements based on the 5W1Hs design space~\cite{lee2023design, shin2023reality} to} gather the common tasks users undertake in the two scenarios, their preferences regarding situated visualization, desired locations for data displays, and the approaches for activating situated visualization in public settings.
Based on the findings, for \textbf{RQ2}, we compiled the appropriate input modalities and proposed a design space with the interaction, required situated visualization and tasks for the public daily scenarios.
We then conducted an iterative design and implementation process with six participants to develop suitable interactive techniques for situated visualization. We distilled key findings from this process, incorporating user feedback.
To investigate \textbf{RQ3}, we created an AR prototype with the designed interaction and tailored it for specific public scenarios referred to as \textit{SituInStore} for the store setting and \textit{SituInLib} for the library. We also provided the baseline for the two scenarios with the same situated visualization but only had hand-based interaction~\cite{elsayed2016blended}.
With these AR prototypes, we conducted a within-subjects field study involving 14 participants to assess the effectiveness and user acceptance of the proposed interaction in public settings. We gathered both quantitative and qualitative feedback through post-study questionnaires and interviews. 
Overall, participants expressed a preference for our prototype compared to the baseline, as they found the interaction to be more acceptable, flexible, and useful in public environments. 

In summary, our key contributions consist of 1) a formative study that understands and explores users' interaction needs of situated visualization within public environments, 2) \zq{ interaction designs with the considerations of users’ awareness of the surrounding environment and the social context,} 
and 3) a field study that evaluates the proposed interactions and the derived design implications regarding appropriate interactions with in-situ data visualizations in public settings.





\section{Related Work}
\subsection{Situated Visualization and Situated Analytics}
\subsubsection{Evolving Situated Visualization}
White et al.~\cite{white2009interaction} proposed the term situated visualization to describe the form of data representation that integrates emerging technologies to display visualizations within their relevant physical and semantic contexts. 
They evaluated the practical advantages of situated visualization by working closely with botanists and urban designers on specific tasks and underscored the potential of situated visualization to enrich people's everyday experiences~\cite{white2009sitelens, white2006virtual}.
Since then, situated visualization has evolved over the past decade and received more attention recently.
ElSayed et al.~\cite{elsayed2016situated, elsayed2016blended} introduced the concept of situated analytics by considering Augmented Reality (AR) and Visual Analytics (VA) domains to facilitate analytical reasoning of data within a physical context.
This concept encompasses not just the contextual display of data, but also the application of analytical methods and decision-making processes within the context~\cite{thomas2018situated}. 
Willett et al.~\cite{willett2016embedded} contributed to the field by proposing a widely cited model, named embedded visualization, which describes the close physical integration between situated data and its physical referent.
They explicitly introduced physical referents for situated visualization and define the relationship between situated \zq{v}isualization and its physical referents. 
Recent work by Satriadi et al.~\cite{satriadi2023proxsituated} expands the model by defining \textit{ProxSituated} and \textit{ProxEmbedded} representations, considering physical proxies and environmental proxies. 
In addition to the models, Bressa et al.~\cite{bressa2021s} provided a survey about situated visualization works, with several situatedness definitons.

\subsubsection{Designing Situated Visualization in Real-world Scenarios}
Given the reality-based nature of situated visualization, designing situated visualization and interaction needs a thoughtful examination of user perception, cognitive processes, and the immediate environmental context~\cite{thomas2018situated, lin2021towards, ens2021grand}. 
Previous research engaged users in workshops and co-design studies to gather ideas for crafting situated visualization~\cite{bressa2019sketching, bressa2022data, jansen2022envisioning, alallah2020exploring}. 
For instance, Bressa et al.~\cite{bressa2019sketching} organized workshops involving ideation and sketching activities to design situated visualizations in everyday situations. 
Some studies have concentrated on tailoring situated visualizations for specific scenarios to assist people's in-situ tasks and decision-making across various domains, such as urban planning, manufacturing, shopping and entertainment~\cite{xu2022arshopping, bach2017drawing, white2009sitelens, becher2022situated, koeman2015everyone, chen2023iball}. 
For instance, 
\zq{Lin et al.}~\cite{lin2021towards} designed an embedded visualization system to assist sports fans in their game-watching experience.
Moreover, Prouzeau et al.~\cite{prouzeau2020corsican} created an authoring tool for situated visualization and demonstrated the potential of involving users in tailoring visualization experiences.
Two recent surveys have comprehensively summarized related works in situated visualization and analytics~\cite{lee2023design, shin2023reality}.

While a substantial body of research has concentrated on the visual representation and presentation of situated visualization~\cite{lee2023design}, or designing interactions on mobile AR devices~\cite{shin2023reality}, less work focuses on interaction design~\cite{elsayed2016blended, satriadi2022tangible}. In addition, existing works mainly leverage the physical context of situated visualization to design interactions, without considering that users may be aware of other people when interacting with situated visualization, especially for public scenarios. 
Our work aims to fill this gap by adopting a user-centered method to explore and design appropriate interactions with situated visualization in public.



\subsection{Design and Evaluate Interaction in Public Environments}
\subsubsection{Designing Interaction in Public}
As technology has progressed, interaction has transcended individual experiences to become a public affair~\cite{reeves2005designing}. 
This paradigm shift presents new challenges for interaction design because it needs to consider a broader perspective that includes not just the dialogue between users and interfaces but also the contextual factors that exert an impact on user interaction~\cite{koelle2020social, rico2010usable}.
Reeves et al.~\cite{reeves2005designing} defined a classification that contains the visibilities of effect and manipulation of interactions \textit{(i.e., secretive, magical, expressive or suspenseful interactions)}, which are frequently linked to the acceptability of interaction used in public. 
The content displayed by AR is personal, but the user's actions are visible \zq{to public}. Therefore, interactions with wearable AR devices often lead to \textit{suspenseful interactions}, \zq{which reveal user manipulations while hiding the interaction effects}~\cite{ens2015candid}.
To provide acceptable interaction in public, it is essential to minimize the effects of interaction on others or provide explanations to avoid socially awkward situations~\cite{ens2015candid, Jung_Philipose_2014}. 

In recent years, there has been a growing body of work investigating the design and evaluation of suitable interaction techniques in public scenarios~\cite{williamson2019planevr, alallah2018performer, schwind2018virtual}.
Several works evaluate the use of existing interactions in public and underscore the importance of understanding user needs early in developing interactions~\cite{montero2010would, ahlstrom2014you}. This process involves considering not only state-of-the-art algorithms but also the contexts in which users agree to use them~\cite{ronkainen2007tap}.
Prior works work investigate the potential usage of AR glasses in conversational scenarios~\cite{koelle2015don}. For example, 
Shardul et al.~\cite{sapkota2021ubiquitous} compared different subtle interaction techniques for AR glasses in everyday settings and Lee et al.~\cite{lee2018designing} explored the design of hand-to-face interaction for AR wearables in public environments.

\begin{figure*}[htbp]
  \centering
  \includegraphics[width=\linewidth]{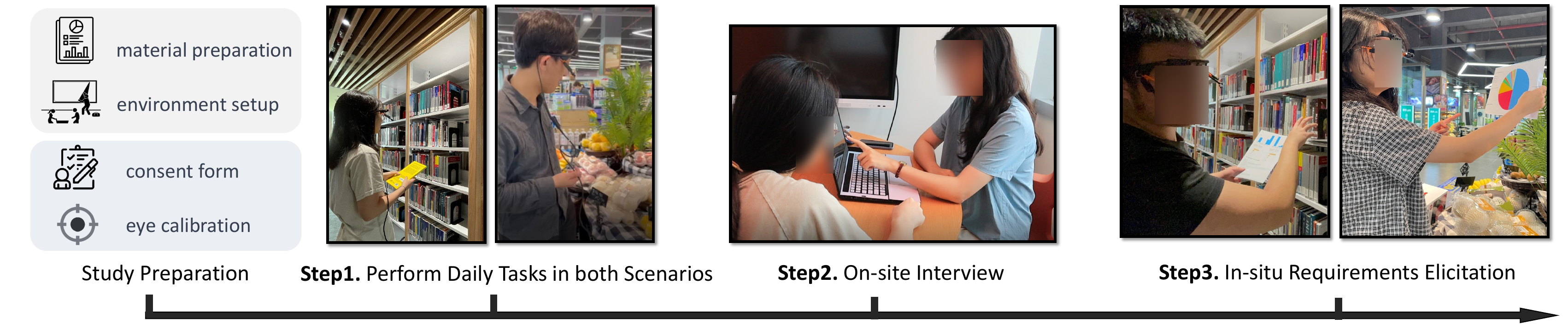}
  \caption{The study comprises three main steps: 1) participants performed the routine tasks, 2) we conducted an interview to collect their activities and requirements, 3) they demonstrated and the detailed requirements of interaction and data display.}
  \label{fig:formativeProcedure}
\end{figure*}

\subsubsection{Evaluating Interaction in Public} \label{relatedwork2.2.2}
Technology acceptance has long been a key aspect of understanding and evaluating the human factors in interactive systems~\cite{koelle2018acceptable, Foltz2008UsabilityEng}. 
Following Goffman's theory from sociology~\cite{goffman2002presentation}, Brewster et al.~\cite{brewster2009gaime} first suggested considering both users and spectators when evaluating the acceptance of emerging technology. 
This duality of user/spectator was described by Montero et al. as social acceptability or social acceptance\footnote{The two terms were used interchangeably by early works~\cite{koelle2018acceptable, brewster2009gaime, montero2010would}.}, which includes two perspectives, user's acceptance and spectator's acceptance~\cite{montero2010would}. 
When measuring social acceptance, prior studies have assessed user acceptance individually by measuring the performer's overall impressions of experiencing the technology or tasks (e.g., feeling comfortable or uncomfortable)~\cite{serrano2014exploring, freeman2014towards, koelle2017all, sapkota2021ubiquitous, cai2023paraglassmenu}. 
Conversely, some research has concentrated on assessing the spectator's acceptance, capturing their perceptions of the performer's actions (e.g., appearing weird or normal)~\cite{profita2013don, koelle2018your, profita2016effect}. In addition, certain works consider both perspectives for a more comprehensive understanding~\cite{ahlstrom2014you, williamson2011send, alallah2018performer, rico2009gestures}. A widely adopted approach is the audience-and-location axes proposed by Rico et al.~\cite{rico2010usable, rico2009gestures}, which considers the duality of social acceptability.

Prior research has investigated the presentation of situated visualization in public~\cite{whitlock2019designing, moere2012designing, jose2014dimensions, white2009sitelens, valkanova2015public}, yet little attention has been paid to the design of interactions and the evaluation of the effectiveness of interactions in public settings.
ElSayed et al.~\cite{elsayed2016blended} designed interactions with situated visualization used in public supermarkets, but they merely leveraged the physical referents or contexts without considering users' awareness of other people within the environment.
Furthermore, it poses unique challenges and considerations of the interaction design for situated visualization due to the diverse data types and associated data analysis tasks. 
Our work aims to fill this gap by first proactively investigating user requirements for situated visualization and interaction in public environments during the early stages.




\subsection{Interaction with AR Visualization in Real-world Scenarios}
The development of immersive technologies expanded the realm of the interaction of visualization, providing a broader space of the interaction design \zq{than using traditional screens}~\cite{lee2012beyond, lee2021post, bach2017immersive, ens2020envisioning}. 
However, it has also brought about a set of challenges, such as the complexity of designing interactions with more than two degrees of freedom (DoF) or exploiting appropriate multimodal interactions~\cite{ens2021grand, buschel2018interaction, elmqvist2011fluid}. 
\zq{AR devices are commonly used as a medium for presenting situated visualization, such as mobile and wearable devices~\cite{becher2022situated, lee2023design, shin2023reality}. Our work draws inspiration from the previous interaction design of visualization presented in wearable AR devices.}


Previous research on AR visualization interaction can be categorized based on the input modalities, including mid-air gestures, tangible interfaces, touch, gaze, speech, spatially-aware objects, and hybrid device or multimodal input~\cite{buschel2018interaction, lee2021post, grubert2023mixed}.
For example, several studies apply mid-air gestures to facilitate pan-and-zoom interactions, map navigation, or the interaction with static AR visualization for real-world tasks or collaboration~\cite{nancel2011mid, satriadi2019augmented, kim2017visar}. 
Researchers have also investigated tangible-based interactions~\cite{billinghurst2008tangible, cordeil2017design, ishii1997tangible, besanccon2016hybrid} by leveraging familiar everyday objects like paper cards, spheres~\cite{bach2017hologram, spindler2010tangible, tong1912exploring}, or easy-to-handle custom items, such as globes and axes~\cite{satriadi2022tangible, cordeil2020embodied} for tangible feedback~\cite{englmeier2020tangible, ssin2019geogate, krug2022clear}.
\zq{These studies demonstrate the advantages of utilizing various interaction inputs for visualization. However, they primarily concentrated on interacting with visualizations presented in AR, with limited consideration given to the visualizations' referents and the broader situated context.}
White et al.~\cite{white2009interaction} explored a range of tangible-based interactions for triggering, searching, and selecting situated visualization. Afterward, ElSayed~\cite{elsayed2016blended} proposed a blended method by fusing situated visualization with physical surfaces, and Satriadi et al.~\cite{satriadi2023active} exploited the interactions for situated visualization by manipulating its physical proxies.
However, \zq{they mainly leveraged the physical referents or contexts of situated visualization when designing interactions}, lacking a comprehensive consideration of the environmental or social context, as well as user acceptance in public. \zq{These factors are crucial for the practical adaptation of interactions in real-world scenarios~\cite{jacob2008reality}.}
\section{Formative Study: Understanding the In-situ Requirements}  \label{formative}
\zq{To capture users' needs and the constraints in public environments,} we first conducted a formative study with users in public environments. 
We chose two typical scenarios that have been used in previous works of situated visualization~\cite{bach2017drawing, xu2022arshopping, elsayed2016blended}, i.e., shopping in stores and book selection in libraries.

\subsection{Participants}
We recruited 12 participants using university mailing lists (6 females, 6 males). They were graduate students or research assistants (ages ranging from 21 to 30, $Mean = 25, SD = 2.78$) in local universities with various backgrounds, including computer science, mathematics, digital media, and communication engineering. They self-reported going to groceries or university libraries once a week on average. All participants were familiar with standard visualization charts, either through their personal or professional experiences. Five participants had prior experience using AR headsets and mobile AR to view information, while the remaining participants had only used mobile AR applications before.

\begin{figure*}[htbp]
  \centering
  \includegraphics[width=\linewidth]{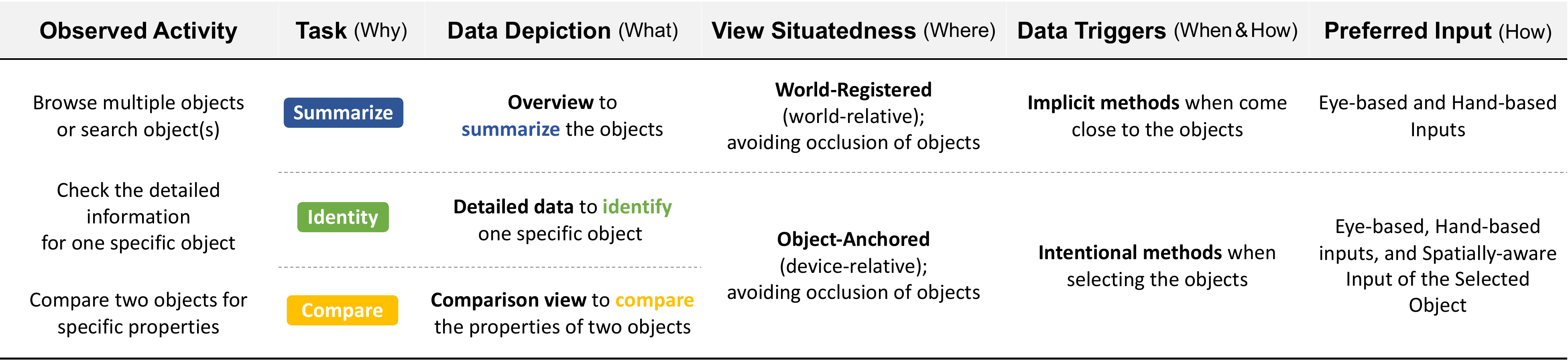}
  \vspace{-2mm}
  \caption{Summary of user requirements for situated visualization and interaction. \zq{Activities, tasks, and data visualizations align with previous works~\cite{lin2022quest, lin2021labeling}, with unique findings about view situatedness, data triggers, and interaction inputs.}}
  \vspace{-3mm}
  \label{fig:findingsFormative}
\end{figure*}
\subsection{Study Setup}
\subsubsection{\textbf{Study Environment, Device and Material Preparation}}
We obtained permission for the two environments \zq{that have varying levels of foot traffic reported by the administrators. This difference can enhance the ecological validity of our study.}
During the study, we utilized a pair of lightweight eye-tracking glasses (i.e., the Pupil Core glass~\cite{kassner2014pupil}) to log users' situated experiences while not impeding their natural movements or activities. The device recorded the participants' first-person perspective behaviors \zq{in real-time videos}, \zq{with overlaid circles representing the focusing areas}.
We leveraged the video recordings from the glasses to help participants recall their perceptions and express their demand for decision support wherever appropriate during post-task interviews. 
Additionally, we printed common visualizations of goods/book data on transparent materials or papers to simulate the display of such data in AR. 

Note that in this work, we mainly focus on designing usable and acceptable interactions among users, physical items within these public spaces, and situated visualizations of these referents. 
Similar to previous studies~\cite{elsayed2016blended, elsayed2015using}, we excluded tasks, such as pathfinding or navigation, in public environments due to safety concerns of the research sites. 

\subsubsection{\textbf{Study Procedure}}  ~\label{procedure}
We obtained participants' consent before the study. Upon arrival at the public research sites (grocery and library in a randomized order for each person), we explained the purpose and procedures of the study. Next, we asked each participant to put on and calibrate the wearable glasses and then proceed to the tasks. 
In each environment, we followed a structured process with three main steps illustrated in Fig.~\ref{fig:formativeProcedure}. The whole study lasted around 55 minutes on average at each location. 
\\
\textbf{Step 1:} 
We invited participants to engage in typical activities within each environment under the following two cases: 1) selecting an item with a specific predetermined purpose, and 2) exploring items placed on designated shelves without a pre-defined objective. These two situations were set to observe and comprehend the decision-making processes in top-down and bottom-up tasks~\cite{park1989product}, as well as the potential needs and requirements for situated data support. We recorded the entire step from a first-person view.
\\
\textbf{Step 2:} 
We moved the participants to a quiet corner of the study sites for a follow-up interview.
We employed a retrospective think-aloud approach~\cite{poole2006eye, lazar2017research}. 
Specifically, we presented the first-person videos captured through the wearable glasses with the participant's gaze visualized and overlaid on top of the camera view~\cite{poole2006eye, lazar2017research}. 
We asked the participants to recall and elaborate on their decision process while watching the videos. In this process, we invited them to specify the possible data types, data representations, displays, and potential interactions that could aid their in-situ tasks or decisions. We took notes and audio-recorded the interviews.
\\
\textbf{Step 3:} 
Following the notes from the interview, we further invited the participants to physically demonstrate how they envision the required data being displayed in the public spaces and the interactions they would like to apply to activate and interact with the situated visualization when there are other people around. 
The participants used printed sheets of visualizations as props to showcase their ideas and preferences.
We recorded their actions, demonstrations, and think-aloud data using the Pupil glasses and an external camera. 

\subsubsection{\textbf{Analysis}}
The first author transcribed the interviews and think-aloud data from the audio and video recordings. Three co-authors collaborated in coding the data using the thematic analysis~\cite{clarke2015thematic}. We generated codes independently, followed by two rounds of discussions to compare, group, and refine the codes, resolving any discrepancies. We carefully reviewed the themes to merge them into categories. 
For the videos captured in Step 3, two authors carefully reviewed the footage, taking note of the participants' preferred positions and interactions of visualization as illustrated.
We filtered out the interaction requirements obtained from the videos proposed by fewer than four participants. 
We concluded with user requirements of situated visualization in the two public environments.

\subsection{Findings} \label{formativeStudyFindings}
All participants reported that situated visualization would effectively assist their routine tasks in both scenarios. 
We adopted the 5W1Hs \textit{(why, what, where, when, who, and how)} design space and the design dimensions of situated visualization~\cite{shin2023reality, lee2023design} to thematically organize and systematically analyze users' needs reflected.
We distilled the specific tasks (why) based on the participants' activities observed and the interviews, along with the pertinent data required on the fly (what). Then, we gathered the desired locations for displaying the data (where), the way (how), and the timing (when) they would like to trigger the situated data. We summarized our findings in Fig~\ref{fig:findingsFormative}. 
As we determined to target ordinary users, we excluded the dimension \textit{Who} in the summary.

\subsubsection{\textbf{Activity $\rightarrow$ Task $\rightarrow$ Data: What Types of Data is Required and Why?}}
We found that participants are engaged in similar activities \zq{and their requirements align with Schneiderman's mantra~\cite{shneiderman1996eyes}} in both scenarios.
When first approaching a display shelf, they tended to initiate by \emph{browsing} all the items and required an \textbf{overview} to summarize the overall categories of objects.
Once locating an interested item, they shifted to the \emph{inspection} activity and hoped to exploit the \textbf{detailed} facet(s) of the object that is not easily accessible offline (e.g., book reviews or sales volume of products)~\cite{xu2022arshopping}.  
If encountering more than one option (mentioned by seven participants), they changed to the pairwise \emph{compare} activity and required a \textbf{comparison view} of the selected objects for easy comparison of properties.
The majority of participants (9/12) emphasized their tendency to compare two objects rather than picking up multiple ones in the given scenario. Therefore, we mainly focus on the pairwise comparison situation.


\subsubsection{\textbf{Data $\rightarrow$ Display: How to Present the Data and Where to Display them?}}
Participants preferred intuitive and familiar data representations for daily tasks, thus we chose standard 2D visualizations, such as bar charts and line charts, displayed in AR (2.5D\zq{~\cite{dwyer2004two}}) and well-known to people~\cite{lee2020reaching}. 
Regarding locations of the visualization, most participants (10/12) would like to place the overview near all objects, using shelves or bookshelves as a reference. As the shelves are generally not moved and people cannot move all the objects at the same time, we take this as a world-registered display. However, the overview is not completely static and is adaptive to the user's location for easy access (world-relative).
Additionally, participants requested having AR category labels based on the overview and linking them to each corresponding object.
For detailed inspection and comparison, they required the visualization(s) to align with the selected object(s) for easy access (object-anchored).
Furthermore, all participants emphasized emphasis on the importance of preventing the occlusion of real-world objects by situated visualizations, as it could obstruct the view of physical objects\zq{~\cite{assor2023handling}}.

\subsubsection{\textbf{Data $\rightarrow$ Interaction: How and When to Trigger the Situated Visualization?}}
We found that the participants preferred various triggers for specific data types and preferred on-demand data display over constant visibility~\cite{lee2023design}.
For the overview corresponded to multiple referents, over half of the participants (8/12) indicated a preference for \textbf{implicit} methods. 
They recommended leveraging implicit ways without explicitly interacting with the objects, such as automatically triggering the chart upon approaching the shelf (proxemics) or browsing through the items.
On the contrary, for activating the detailed view with one referent, nine of them expressed the need for \textbf{controlled methods} to indicate their intention of accessing information. This is to avoid \textit{``having all the details presented at once''} (P2) as it {``can be overwhelming and may lead to occlusion issues.''} (P8).
They expressed similar requirements for activating the comparison view (two referents). Over half of the participants suggested leveraging behavioral cues such as \textit{``natural putting the items close together''} (P4), which aligns with the prior work~\cite{elsayed2016blended}.

\subsubsection{\textbf{Context $\rightarrow$ Interaction: How to Interact with In-Situ Data in a Publicly Acceptable Manner?}}
In both public settings, the participants ruled out voice input. They noted that \textit{``speaking in the library may disturb others''} (P4), and found that it could be \textit{``too noisy to effectively use voice commands in the supermarket.''} (P9). 
Recent research in HCI has started exploring the voice interaction used in noisy or public environments (e.g., silent speech)~\cite{chen2021understanding, yan2019privatetalk}, but the technologies are still in the developmental stages and thus we leave it for the future work.
\zq{All participants expressed a preference for interaction involving hand-based input for the required tasks. 
Notably, 10 participants highlighted their preference for eye-based input to achieve subtle interaction effects, deeming eye-based input as a convenient, natural, and socially acceptable approach in public settings. 
Specifically, for the overview, they indicated a willingness to use both eyes and hands and for the comparison view or detailed view, they also considered manipulating handheld objects as a means of interaction.}
In addition, seven participants refrained from employing expansive arm movements or mid-air gestures in both environments. These actions could be interpreted as \textit{``weird by others''} (P10) when performed publicly. Instead, they tended to gravitate towards minimal and inconspicuous physical actions within their personal space. 
In other words, people's willingness to engage in an interaction is substantially influenced by their immediate surroundings and the presence of other individuals. 

\subsection{Summary}
Our formative study results confirm the demand for situated visualization and identify the need for interaction in public environments. Generally, the participants exhibited distinct interaction preferences for different kinds of visualization, with contextual factors greatly influencing their preferences. 
\zq{However, relying on static printed visualizations is insufficient} for effectively eliciting nuanced design considerations or the subsequent interactions (e.g., filtering) after triggering the situated visualization.
We delve into this exploration in the subsequent iterative design process.
\begin{figure*}[htbp]
  \centering
  \includegraphics[width=0.96\linewidth]{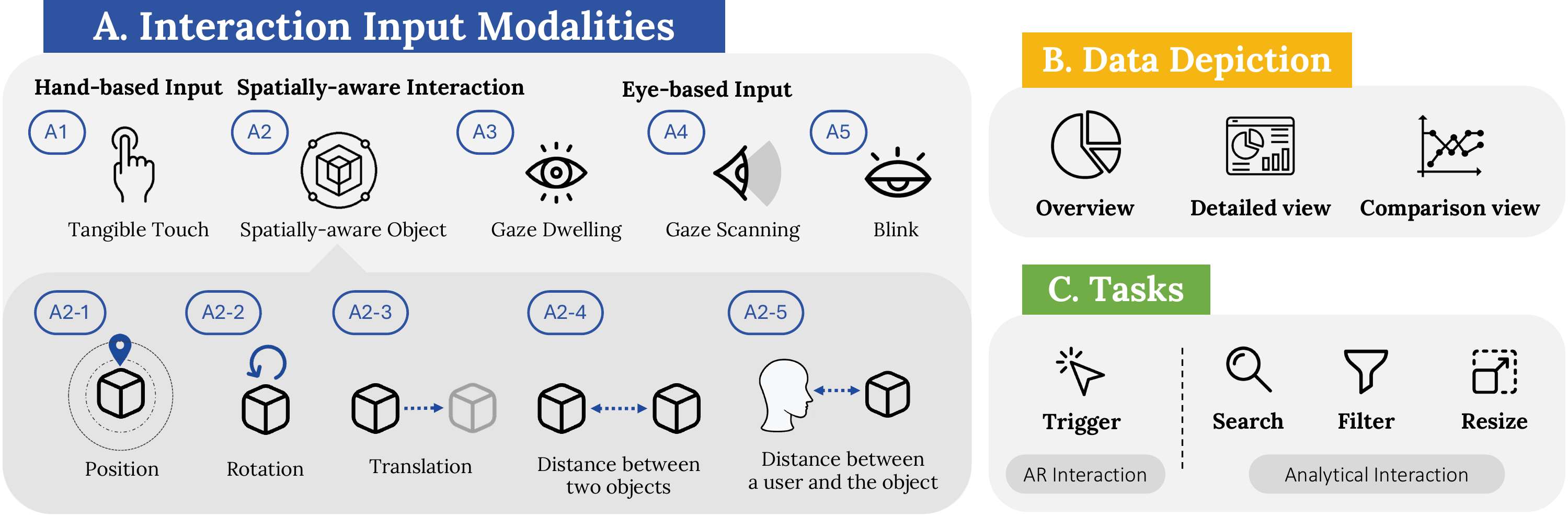}
  \vspace{-2mm}
  \caption{The framework for interaction design exploration within the chosen scenarios. In (A), we present interaction input modalities for selected public scenarios and the required data views in (B), informed by the formative study. We summarize \zq{the required tasks} with situated visualization, detailed in (C), identified through iterative design.}
  \vspace{-3mm}
  \label{fig:DesignSpace}
\end{figure*}
\section{Interaction Design Exploration} \label{Iterative4DesignSpace}
\begin{figure*}[htbp]
  \centering
  \includegraphics[width=0.9\linewidth]{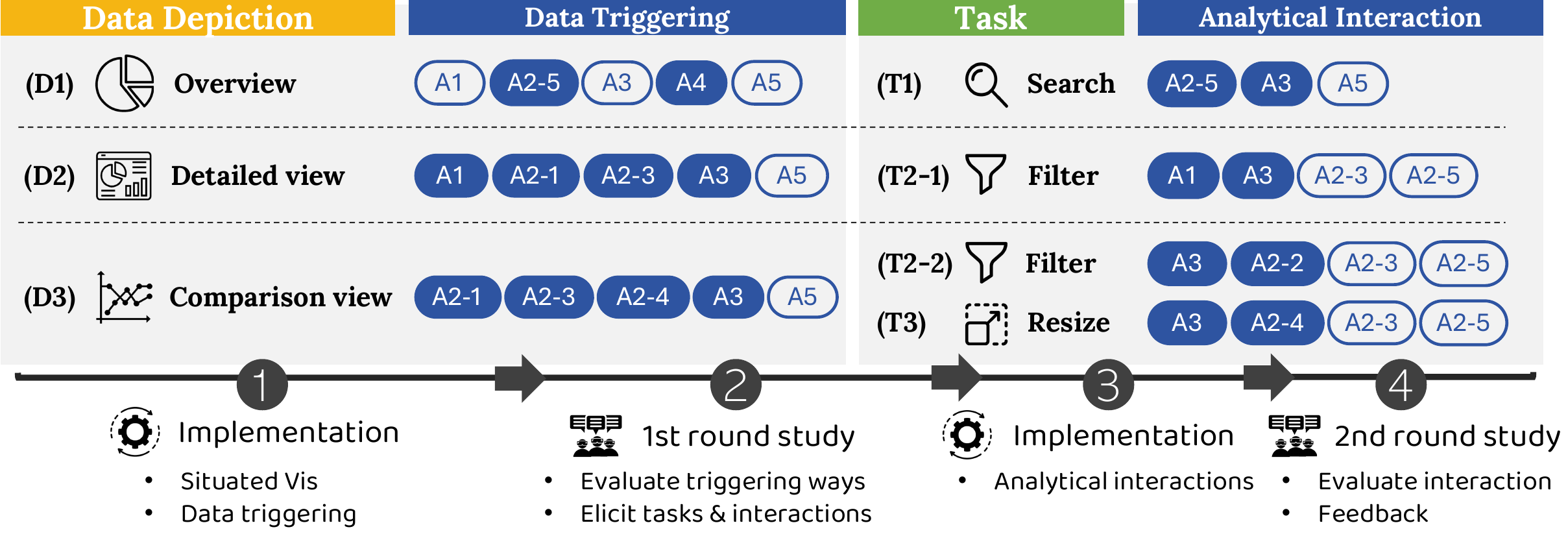}
  \vspace{-2mm}
  \caption{The iterative design process with four steps that alternate between the implementation and design sessions. 
  Based on the formative study's findings, we implemented the initial data triggering interaction with situated visualization (Step 1). Then, we conducted two rounds (Step 2-4) of design study with the participants to explore, evaluate, and improve the interaction.}
  \vspace{-4mm}
  \label{fig:IterativeDesign}
\end{figure*}

To address \textbf{RQ2} of exploring the appropriate interaction techniques for situated visualization in public, we first distilled a range of interaction input modalities in AR (Fig.~\ref{fig:DesignSpace}) that can be acceptable to users and suitable for public contexts informed by the formative study and the literature.
Then, we organized an iterative design process (Fig.~\ref{fig:IterativeDesign}) with six participants to design, actualize, and improve the interaction techniques. 

\subsection{Interaction Input Modalities}
The input modalities (Fig.~\ref{fig:DesignSpace} A) comprise a spectrum of AR interactions based on the required situated visualization in public. We mainly considered hand-based, eye-based and \zq{spatially-aware interactions\footnote{We adopt the term \textit{spatially-aware interaction} as a scaleable method referring to ~\cite{kortuem2009smart, ye2023proobjar}. In this paper, it specifically denotes the use of spatially-aware non-digital objects for the selected scenarios.}} as plausible modalities. 
\begin{itemize}
  \item For the hand-based interaction, we incorporate the tangible touch on the physical objects that serve as the referents for situated visualization (Fig.~\ref{fig:DesignSpace} A1) as it is a natural and basic action people perform in real life.
  We use the term \emph{tangible} over \emph{tactile} or \emph{haptic}, as our emphasis lies in leveraging the tangible feedback from objects, rather than relying on tactile or haptic mechanisms to convey information~\cite{shaer2010tangible}. 
  
  \item We selected the spatially-aware object interaction in Fig.~\ref{fig:DesignSpace} A2 based on people's activity observation in the formative study. They usually pick up objects of interest and examine the detailed information in their hands. In addition, spatially-aware object interaction has been widely applied in various daily scenarios and used as the interaction medium for AR visualization~\cite{radle2015spatially, elmqvist2013ubiquitous, langner2021marvis}.
  We further divide the interaction into five subcategories, namely object position relative to a designated area (A2-1) and relative position to the user (A2-5), rotation (A2-2), translation (A2-3), and the distance between two distinct objects (A2-4). We exclude dynamic motion (e.g., speed or moving patterns) and complex spatial relationships (e.g., relative angles) from our considerations, recognizing the constraints for precise motor control (physical) and sensing (technological). 
    \item Regarding the eye-based input, we identified gaze dwelling (A3), gaze scanning (A4), and eye blinking (A5) which are common and natural signals of human attention or implicit intention~\cite{holmqvist2011eye} and can be captured by AR head-worn devices (e.g., Hololens 2, Magic Leap 2).
    Gaze dwelling conveys the desire to learn more about the entity at one's focal attention~\cite{penkar2012designing}. 
    Gaze scanning is a behavior of quickly collecting information about a set of items or across an area~\cite{biswas2013new}.
    Intentional eye blinking, on the other hand, can play the role of a subtle communicative signal for humans to imply their interest in continuing the interaction~\cite{hori2006development}. Many AR applications have employed these eye-based inputs to trigger or manipulate the virtual information in daily life~\cite{lu2020glanceable, hirzle2019design, duchowski2018gaze}. 
    In this work, we consider these eye-based input methods to help users convey implicit intention and capture their unconscious natural behavior and attention.
\end{itemize}
The actual choice of interaction modality is influenced by the \emph{data} involved (Fig.~\ref{fig:DesignSpace} B) and the analytical \emph{tasks} (Fig.~\ref{fig:DesignSpace} B) to perform. We explore the appropriate interaction techniques through an iterative design process with users.

\subsection{Iterative Design for Data Triggering and Analytical Interaction} \label{iterative1stround}
We extended invitations to all participants of the formative study to engage in the design process; six of them agreed. We then organized an iterative design process, as shown in Fig.~\ref{fig:IterativeDesign}, which alternated between technical implementation and design feedback sessions with the six participants.

\subsubsection{Designing Interaction for Triggering Situated Visualization  (Step 1-2)}
According to the formative study and the literature~\cite{shin2023reality}, triggering situated visualization is the events that lead to the instantiation. We approach this by first implementing various interaction techniques with the required visualization and then inviting users to experience them to select the suitable choice(s).
We determined the data presentation and display of the situated visualization in AR based on the findings in Sec.~\ref{formativeStudyFindings} (Details see Sec.~\ref{prototype}).
The initial AR prototype with registered situated visualization and interaction techniques was crafted using Microsoft's Mixed Reality Toolkit (MRTK) on HoloLens 2, leveraging Vuforia target tracking for object recognition and tracking~\cite{vuforia2023}.


For triggering the overview in Fig.~\ref{fig:IterativeDesign} D1, we incorporated suitable input modalities (Fig.~\ref{fig:DesignSpace} A) that consistent with the findings (Sec.~\ref{fig:findingsFormative}). 
We explored three gaze-based interactions: gaze dwelling (A3), gaze scanning (A4), and eye blinking (A5). These interactions are enabled as triggers only when the user is oriented toward the designated set of target objects at a sufficiently close distance (A2-5).
Additionally, we introduced a tangible touch method (A1) as an alternative of activating the overview. This approach involves users physically touching a specific area overlaid on the shelf containing the desired objects, rendering it natural and acceptable in public.

To deliberately trigger detailed data (Fig.~\ref{fig:IterativeDesign} D2) of an object, we introduced a variety of options that allow for intentional control of data while seamlessly integrating with users' customary behaviors. 
These options included tangible touch (A1), spatially-aware object (A2), gaze dwelling (A3), and eye blinking (A5). The two eye-based methods shared similar mechanisms for activating the overview. 
Our implementation of tangible touch involved the placement of a pressable button overlay on the object, which could be conveniently accessed through natural touch or clicking gestures~\cite{elsayed2016blended}. 
We also experimented with manipulating the object's position (A2-1) and translation (A2-3) to activate the detailed data, i.e., moving the object into a designated virtual region around referring to previous work~\cite{white2009interaction}.

To initiate the comparison view between two objects (Fig.~\ref{fig:IterativeDesign} D3), we leveraged proximity (A2-4) as an indicator of the user's intention to engage in a pairwise comparison, aligning with our everyday behaviors~\cite{elsayed2016blended}. 
Consequently, when two objects are detected in close proximity within the user's field of view, the corresponding data properties shared between them become available. To select a specific property for comparison, we employed the same triggering methods (i.e., A3, A5, A2-1, and A2-3) used for accessing detailed data when multiple data properties are available. However, tangible touch (A1) is not applicable in this case due to both hands being occupied by holding the two objects.

Next, we conducted the first round of study involving six participants (three female, with sessions lasting 50-70 minutes each) to assess the usability, utility, and user acceptance of these data-triggering techniques. 
The study took place within a laboratory environment, featuring prearranged products and books.
For each type of data in Fig.~\ref{fig:DesignSpace} B, we invited the participants to wear the AR headset to experience the interaction techniques in an embodied way and give feedback~\cite{schleicher2010bodystorming}.
Based on their input, we identified interactions that were perceived as suitable for triggering situated visualization, denoted by solid labels filled with blue color adjacent to Fig.~\ref{fig:IterativeDesign} D1-D3.
These interactions were reserved for further development and field study evaluation, while the interactions with hollow labels in unfilled color were removed considering user acceptance and technical stability after the study. We report the reasons in Sec.~\ref{iteractiveFindings}.


\subsubsection{Designing Interaction for Situated Analytics (Step 3-4)}
During the first study, while the participants had the headset on, we asked about their needs for analytical interactions with the activated situated visualization. We used the findings to guide the second design iteration. 
More specifically, as illustrated in Fig.~\ref{fig:IterativeDesign} T1-T3, users demanded searching for corresponding items based on certain attributes of the overview, filtering data on the detailed and comparative view, and resizing the entire comparative view for a closer look.
Informed by users' verbal feedback and their demonstration of how they wished to analyze the situated visualization in AR, we designed and implemented various interaction techniques for these analytics tasks (Fig.~\ref{fig:DesignSpace} C). 
We considered the technical robustness, alignment with the selected methods used for triggering situated visualization, and perceived naturalness in our design proposals. 
We implemented the interactive techniques and conducted the second study involving the participants to validate the interactions and give feedback to improve them (each session between 30-50 minutes). After the study, we improved the interaction techniques by incorporating visual cues and triggered effects.


\subsection{Key Insights} \label{iteractiveFindings}
The design process enabled us to iteratively explore, develop, test, and validate the appropriate interactions through collaboration with users. It yielded valuable insights into the interaction design in public.

\subsubsection{\textbf{Consistency between Interaction and Task Behavior.}} 
We discovered that the participants prefer turning habitual behaviors performed to satisfy their information needs into controls of situated visualizations for the same purpose.
For instance, three participants indicated that using their sweeping gaze across objects of interest to trigger the overview was natural and in line with their intent compared to eye blinking or gaze dwelling (keeping A4 for Fig.~\ref{fig:IterativeDesign} D1). 
They considered using gaze dwelling to trigger the detailed data of a single object, as it conveys the act of focusing attention on a selected item (reserving A3 for Fig.~\ref{fig:IterativeDesign} D2-D3).
Similarly, participants thought it was common to touch one item at a time to view the situated data for inspection of details (saving A1 for Fig.~\ref{fig:IterativeDesign} D2), while touching the shelf or multiple items consecutively to access an overview was less socially desirable (removing A1 from Fig.~\ref{fig:IterativeDesign} D1). 

\begin{figure*}[tp]
  \centering
  \includegraphics[width=0.99\linewidth]{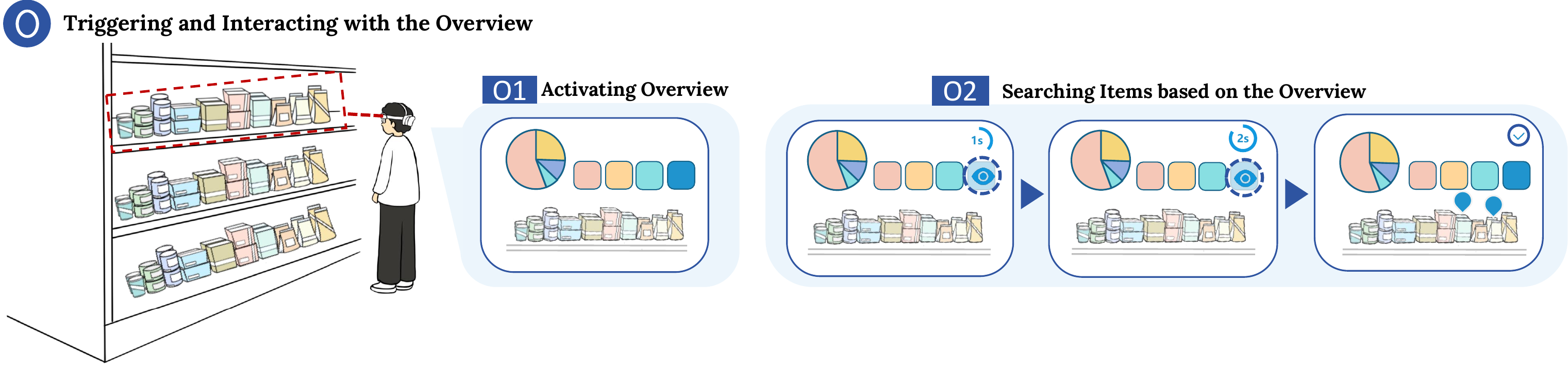}
  \vspace{-2mm}
  \caption{Interaction techniques with the overview. We offer the eye-based scanning interaction to trigger the overview (O1) and the eye-gaze dwelling for searching items with the overview (O2).}
  \vspace{-3mm}
  \label{fig:Overview}
\end{figure*}
\begin{figure*}[tp]
  \centering
  \includegraphics[width=0.99\linewidth]{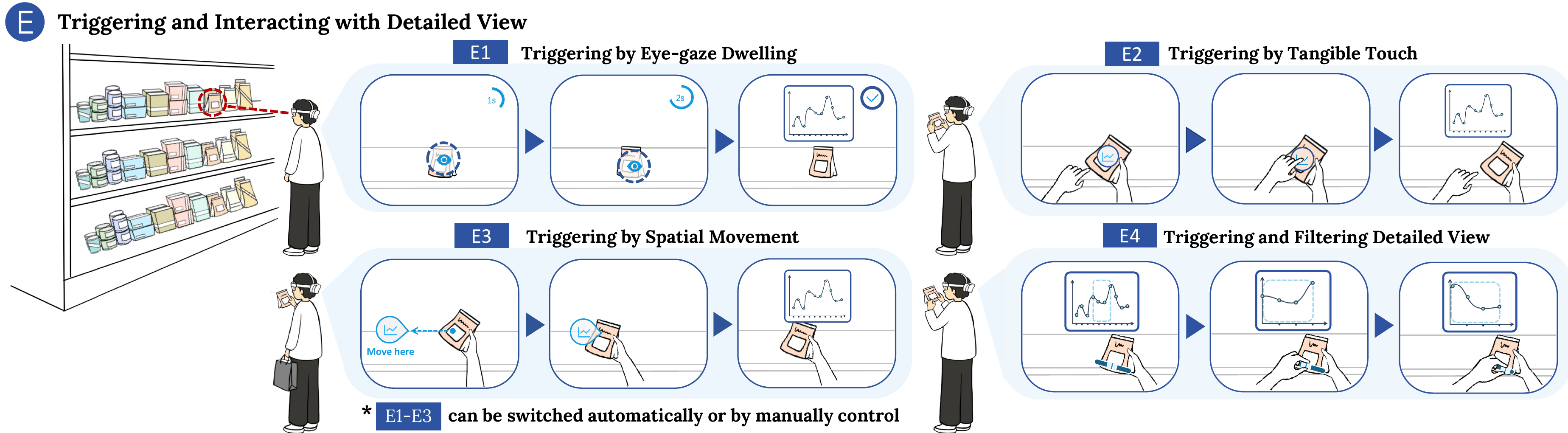}
  \vspace{-2mm}
  \caption{Interactive techniques with the situated visualization that presents the detailed data of one single object. We provide three kinds of techniques for activating the detailed data in (E1-E3) and one technique for selecting and filtering data (E4).}
  \vspace{-3mm}
  \label{fig:Detail}
\end{figure*}

\subsubsection{\textbf{Subtle and Natural Movement.}}
Participants exhibited a preference for interacting through subtle movements that felt organic and akin to their everyday actions. 
For example, they avoided engaging in conspicuous or extensive spatial manipulations of physical objects, recognizing that such actions could potentially disrupt others and draw attention in public spaces (A2-5 excluded from Fig.~\ref{fig:IterativeDesign} T2-1 \& T2-2).
Furthermore, the participants expressed a desire to avoid performing uncommon actions, which could attract unwarranted attention or make them appear weird. For instance, we implemented an interactive technique that requires users adjusting the proximity of an object to \zq{scale} the visualization, \zq{as it was required by half of the participants} (A2-5 in Fig.~\ref{fig:IterativeDesign} T3). Although this interaction aligns with people's natural intent -- pulling something nearer to get a close-up look, \zq{after trying this out with an AR headset on,} participants changed their opinions and reported that they might avoid using it in public for fear that repeating it too often would make others think they had potential vision problems.


\subsubsection{\textbf{Technical limitation of AR headsets.}}
We found that the technical issues of AR head-worn devices can also affect the interaction experience. 
For example, participants expressed confusion regarding the eye-blink interaction because sometimes it was challenging to discern whether blinking was intentional or not. Blinking (A5) was thus removed from all interactions. 
In addition, due to technical limitations, it is difficult to achieve precise filtering with situated visualization, either by hand or by eye. For example, although we could obtain precise object rotation angles through the headset, they cannot be stably sensed and tracked for precise control of data. 
To address these challenges, we proposed two solutions. 
First, for less reliable movements, we only used action tendency (e.g., the act of rotating, A2-2 for Fig.~\ref{fig:IterativeDesign} T2-2) rather than the precise level of movement (e.g., the actual angle rotated) as coarse control signals. 
Second, we integrated familiar widgets for visualization operations (i.e., sliders) and placed them in easy-to-reach-and-calibrate regions (e.g., edge-based~\cite{he2023ubi}) on the corresponding object(s). We tuned the parameters of the widgets through technical testing so that users could exercise a tangible touch on them to achieve relatively fine-grained visualization manipulation (A1 in Fig.~\ref{fig:IterativeDesign} T2-1). 
This approach was designed to enhance the precision and usability of the interaction within the constrained technical conditions, improving the overall user experience.

\begin{figure*}[htbp]
  \centering
  \includegraphics[width=0.99\linewidth]{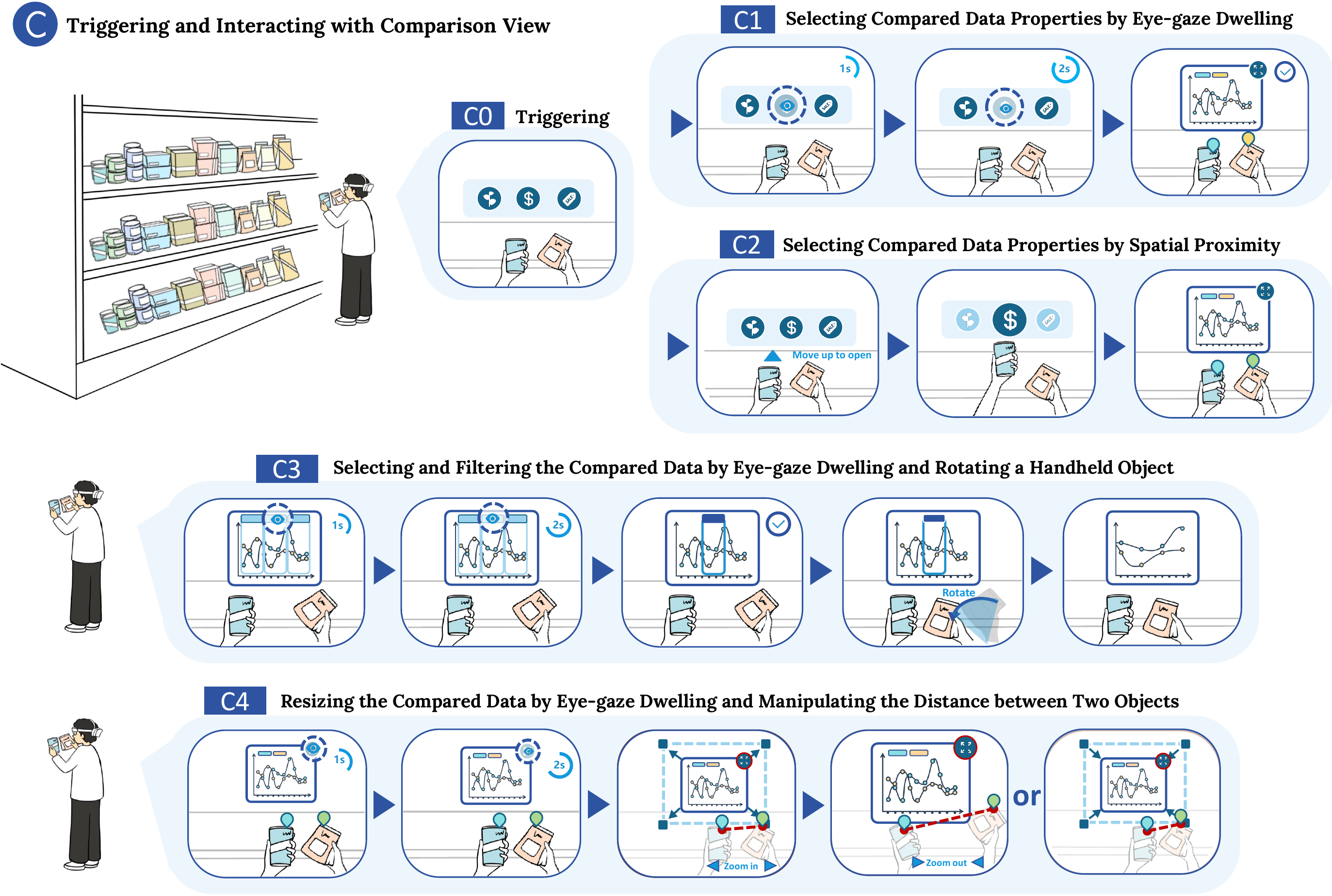}
  \vspace{-2mm}
  \caption{Interactive techniques with the comparison data that presents the aggregated data view of two selected objects. We implement two techniques for activating and selecting the data properties for comparison (C1-C2) and two analytical interactions for filtering and resizing the compared data (C2-C4).}
  \vspace{-4mm}
  \label{fig:Compare}
\end{figure*}
\section{Interaction Techniques with Situated Visualization} \label{prototype}
Building upon the design exploration in Sec.~\ref{Iterative4DesignSpace}, we introduce the final interaction techniques with the situated visualization required by users.
In this section, we illustrated the techniques within the grocery store as examples. We have also customized these techniques for the books in the library by using different data and adjusting the parameters of these techniques based on the library context (e.g., distance, button locations).
A video is provided in the supplementary.

\subsection{Data Collection, Representation and Display}
In both of the chosen public scenarios, we gathered three types of data that were required by a majority of participants (Sec.~\ref{formative}). In the grocery, we gathered the nutrient composition, price, and sales fluctuations that are not readily available offline for products. With the library system and online platforms (e.g., \textit{Amazon}), we collected data pertaining to book reviews, sales, and borrowing quantity.
To present the data, we utilized conventional visualization, including pie charts, line charts, and bar charts, in conjunction with textual information. 
Drawing upon the design dimensions of situated visualization~\cite{lee2023design} and findings in Sec.~\ref{formativeStudyFindings}, we spatially displayed the situated visualizations considering their \emph{situatedness} with referents, \emph{cardinality} and \emph{visibility}. 
All the visualizations were situated rather than embedded to avoid occlusion. They shared the spatial and view coordinate frames with the referents to enhance the semantics connection. Furthermore, we adjusted the orientation of the visualizations to align with the AR camera's frame when activated, ensuring text legibility and ease of viewing.

\subsection{Interaction with Overview} \label{overview}
As depicted in Fig.~\ref{fig:Overview}, we leveraged the eye-based scanning (Fig.~\ref{fig:DesignSpace} A4) to trigger the overview of multiple objects.
When the user approaches the objects within a defined distance threshold (between 0.8 and 2 meters, adjusted according to real-world environments) and scans across them, an overview represented by a pie chart is shown appropriately and could update based on the object being viewed. This pie chart provides an overview of the object categories.
With the overview in view, users can locate specific items by gazing at the relevant legend adjacent to the pie chart. 
To mitigate the Midas touch~\cite{jacob1993eye}, we leveraged the legends rather than the sectors associated with the pie chart and implemented a 3-second gaze dwelling for triggering~\cite{hubenschmid2021stream}. 
We also added an instantly-visible loading indicator as the visual cue that appears with the AR cursor and changes to green once the selection process is complete. Once activated, items belonging to the category represented by the legend are highlighted in situ with AR arrows.

\subsection{Interaction with Detailed Data} \label{detailed}
We ultimately chose three interactive techniques for triggering detailed data for a single object because they were all positively received by the participants, and none stood out as significantly better than the others (Sec.~\ref{iterative1stround}).
As shown in Fig.~\ref{fig:Detail} E1-E3, the user could trigger the detailed data by gaze dwelling (E1), tangible touching on a button overlaid (E2), and moving the object to a designated position (E3), which is indicated by visual cues.

To facilitate a seamless transition between E1-E3, we implemented a mixed-initiative approach for users to switch between them~\cite{findlater2004comparison}. This approach encompasses an adaptive mode as the default setting, along with a toggle menu that allows users to manually select their preferred interaction.
The adaptive mode considers the availability of users' hands and the position of users. When the distance between the user and the object exceeds a predefined threshold (approximately arm's length), gaze dwelling (E1) is recommended. 
when at least one user's hand is tracked in AR and the distance between users and the object falls below the threshold, the other two hand-based methods (E2, E3) are suggested as gaze-dwelling can be unstable when getting close. 
The spatial movement (E3) is recommended when only one hand is tracked within a specific timeframe, considering the difficulty of holding and touching a button with only one hand available.
Users could also preset or manually set a default option using the toggle menu.
The parameters of adaptation were tested and determined during the iterative process.


After activating the data, users could filter the visualization (e.g., line chart) through manipulation of an edge-based opportunistic slider widget, which is automatically aligned with the tracked object's edge~\cite{he2023ubi, henderson2008opportunistic}. 
The slider will appear after triggering a specific visualization. 
Given that multiple visualizations can be activated simultaneously, the slider's functionality will automatically align with the visualization that the user has focused on within three seconds. We introduced this interaction for several reasons: 1) users are accustomed to utilizing sliders for data filtering, enhancing familiarity; 2) a tangible edge-based slider offers a precise and natural means of filtering data; and 3) Edges represent one of the most prevalent geometric features of everyday physical objects, making them an intuitive choice for interaction.


\subsection{Interaction with Comparison View} \label{comparison}
We offered two methods to initiate the comparison view between two objects. When users bring two objects into close proximity (Fig.~\ref{fig:Compare} C0), the data property buttons appear for selection. 
Users could select by gaze dwelling (Fig.~\ref{fig:Compare} C1) or by moving one object in spatial proximity to the button (Fig.~\ref{fig:Compare} C2).
Given that users' hands are occupied in this situation, we relied on hand- and eye-gaze-based input for subsequent analytical interactions. 
Users needed to first activate the relevant buttons using gaze dwelling to communicate their intent and then utilize hand-based operations to confirm their intention. 
To filter the compared data (Fig.~\ref{fig:Compare} C3), the user first activates the area of interest in the visualization by gaze dwelling and then slightly rotates either hand with the object to confirm the data filtering.
To resize the comparison view (Fig.~\ref{fig:Compare} C4), the user triggers the resize button by gaze dwelling and then scales the entire visualization by adjusting the distance between the two objects.


\begin{table}
\centering
\caption{The provided questions for participants to experience the interaction techniques. which were adjusted accordingly to the actual environment and conditions.}
\label{tab:tasks}
\arrayrulecolor[rgb]{0.616,0.616,0.616}
\begin{tabular}
{>{\centering\hspace{0pt}}m{0.054\linewidth}>{\centering\hspace{0pt}}m{0.333\linewidth}>{\centering\arraybackslash\hspace{0pt}}m{0.5\linewidth}} 
\arrayrulecolor{black}\toprule
\multicolumn{1}{>{\hspace{0pt}}m{0.06\linewidth}}
{\textbf{No.}} & \textbf{Interaction (Sec.5)} & \textbf{Question}                                                \\ 
\toprule
1                                                                & Inspect details (\textbf{\textcolor[rgb]{0.145,0.388,0.878}{E1-E3}})          & What is the detailed \textit{information*} of the object?        \\ 
\arrayrulecolor[rgb]{0.616,0.616,0.616}\cmidrule(lr){1-1}\cmidrule(lr){2-2}\cmidrule(lr){3-3}
2                                                                & Search for categories (\textbf{\textcolor[rgb]{0.145,0.388,0.878}{O1-O2}})    & Which products/books fall into the  \textit{given category}?~    \\ 
\cmidrule(lr){1-1}\cmidrule(lr){2-2}\cmidrule(lr){3-3}
3                                                                & Compare information (\textbf{\textcolor[rgb]{0.145,0.388,0.878}{C1-C2}})      & Which object has the higher value of the \textit{information*}?  \\ 
\cmidrule(lr){1-1}\cmidrule(lr){2-2}\cmidrule(lr){3-3}
4                                                                & Filter data of one object (\textbf{\textcolor[rgb]{0.145,0.388,0.878}{E4}})  & What is the concrete value of~\textit{information*} in the past day?   \\ 
\cmidrule(lr){1-1}\cmidrule(lr){2-2}\cmidrule(lr){3-3}
5                                                                & Filter compared data (\textbf{\textcolor[rgb]{0.145,0.388,0.878}{C3}})       & Which one has the higher value regarding the \textit{information*}?   \\ 
\cmidrule(lr){1-1}\cmidrule(lr){2-2}\cmidrule(lr){3-3}
6                                                                & Resize the comparison view (\textbf{\textcolor[rgb]{0.145,0.388,0.878}{C4}}) & Could you resize the compared visualization to view it clearly?         \\ 
\arrayrulecolor{black}\cmidrule[\heavyrulewidth]{1-1}\cmidrule[\heavyrulewidth]{2-3}
\multicolumn{3}{>{\hspace{0pt}}m{0.941\linewidth}}{\textit{\small{*It can be replaced by a product's sales, price change or detailed nutrition information in the store, or a book's online reviews, sales, and borrowing quantity in the library.}}}     
  \vspace{-4mm}   
\end{tabular}
\end{table}
\section{User Study}
To answer \textbf{RQ3}, we conducted a within-subjects study in public scenarios compared with a baseline condition to 1) evaluate the usability, usefulness, user acceptance, and engagement with the interaction techniques; and 2) understand the benefits and limitations of the designed interaction.

\subsection{Participants and Experimental Design}
\subsubsection{Participants}
We obtained institutional IRB approval before recruiting 14 participants (P1-P14; seven females) through the university mailing list.
They, aged 23 to 31, had not previously taken part in the formative study or the iterative design phase. All the participants go to the grocery store or the library at least once a week. Seven of them reported that they had never worn an AR headset before, while four had tried it once or twice and the remaining three were familiar with the AR headset.
\begin{figure*}[htbp]
  \centering
  \includegraphics[width=0.99\linewidth]{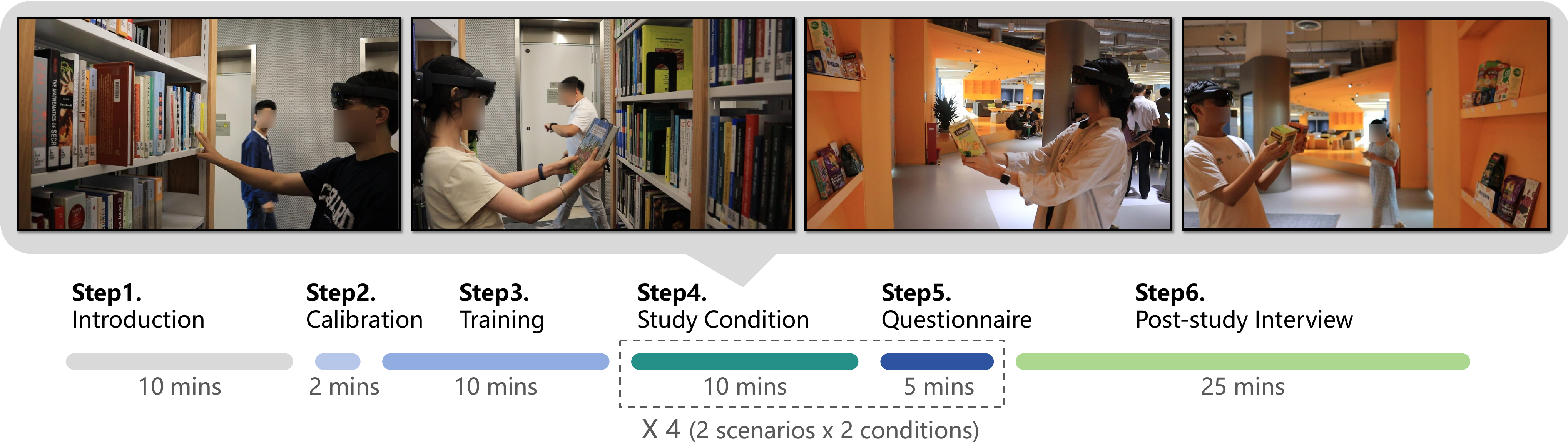}
  \vspace{-2mm}
  \caption{The detailed procedure of our experiment and the photos of study conditions in Step 4.}
  \label{fig:studyProcedure}
  \vspace{-4mm}
\end{figure*}
\subsubsection{Experimental Setup}
We obtained permission to conduct the main experiment in the library but not in the grocery store due to concerns about daily operations and safety.
We thus applied for a position in the campus's public learning center and simulated the setup of a grocery store there, including placing commodity walls and pasting price labels. We prepared 12 books borrowed from the library and 12 snacks bought from the store as study materials. They are displayed in a random order with at least eight items provided each time for the participants to minimize the learning effect.
We pre-loaded the visual images of the books and products into the baseline and our systems and updated the corresponding situated visualization data before each study session.
In all conditions, the participants experienced the interaction by wearing the Hololens 2 AR headset in front of the bookshelf/commodity wall of prepared objects (Fig.~\ref{fig:studyProcedure}). 

\subsubsection{Baseline Condition}
We built a baseline as the control condition to evaluate whether the proposed interaction design is more usable and acceptable to users than the conventional AR interactions in public settings. 
We implemented the baseline by replacing \zq{our proposed new} interactions triggered by gaze or spatially-aware objects \zq{with hand-pointing interactions (touch and/or mid-air) referring to the prior work~\cite{elsayed2016blended}. More specifically, if the target content of interaction was overlaid on a physical object, it was categorized as touch. If the target has no apparent attachment to an object, it is considered a mid-air gesture (e.g., search items in Fig.~\ref{fig:Overview}). For other hand-pointing interactions, the baseline is consistent with our prototype (see the comparison table in the supplementary materials). 
}
All the situated visualization and data displays in the baseline remained the same as in our prototype.
As we had customized versions of proposed AR prototypes for the two selected scenarios (the library and store) with different datasets, we also developed two corresponding baseline counterparts for these settings. These were referred to as \emph{BaselineStore} and \emph{BaselineLib}.

\subsubsection{Rountine Tasks for Experiencing the Interaction}
To ensure a comprehensive user experience involving various interactions in both our AR prototype and the baseline, we devised a series of everyday tasks under the two public decision settings (i.e., library and store) based on the insights gleaned from our formative study and design iterations. 
As detailed in Table~\ref{tab:tasks}, we assigned each participant five specific tasks, in the form of questions. These tasks maintain consistency between our prototypes and baselines across two common scenarios.
Within the same scenario, we ensured that over 50\% of the objects in the two conditions were different for each participant to prevent any learning effect.
These tasks were structured to allow participants to sufficiently engage with the interactions in typical daily scenarios. We did not aim to assess user performance or measure task completion times for comparison during the study.

\subsection{Procedures and Measures}
\subsubsection{Procedure}
Two experimenters conducted the study on the experimental sites, following a structured procedure consisting of six phases as illustrated in Fig.~\ref{fig:studyProcedure}.
We obtained written consent from each participant prior to the study.
\begin{enumerate}
    \item \textbf{Introduction:} We introduced them to the study and provided information about the interactive techniques through first-person tutorial videos recorded using the AR headset.
    \item \textbf{Eye Calibration:} Participants were invited to wear the AR headset (Hololens 2) and calibrate the embedded eye tracker.
    \item \textbf{Training:} A training session was conducted for each participant to acquaint them with using the AR headset and to familiarize them with the interactions presented in both our AR prototype and the baseline.
    \item \textbf{Study Condition:} Participants were assigned to one of the conditions and were presented with specific tasks (see Table~\ref{tab:tasks}). They were required to interact with the situated visualization in AR to find the answers to these tasks. No strict time limits were imposed, but participants were encouraged to seek answers promptly.
    \item \textbf{Questionnaire:} After completing the tasks, participants filled out a questionnaire regarding their experience with the interactions in the condition they had just experienced.
    \item \textbf{Interview:} Upon completing all four study sessions (comprising two public scenarios and two conditions, our prototypes and the baselines), each participant joined a post-study semi-structured interview. These interviews aimed to uncover the reasons behind their questionnaire responses and to explore their subjective feelings regarding interacting with situated data in a public setting. During the interviews, we showed participants videos of their interactions recorded from a third-person perspective, helping them recall and evaluate their actions in a public setting from an external observer's viewpoint.
\end{enumerate}
The order of the four sessions was randomized and the participants \zq{were not informed about the baseline condition during the study.} 
During \textit{Step 4}, we automatically logged the participants' interactions in AR by scripts and one experimenter manually counted the number of passers-by who walked near them during each study session. 

\begin{figure*}[htbp]
  \centering
  \includegraphics[width=\linewidth]{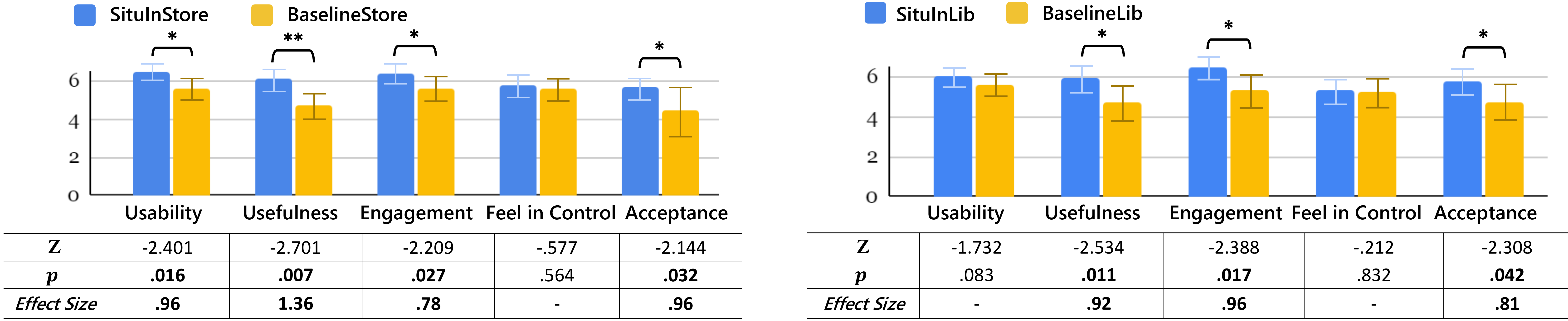}
  \vspace{-2mm}
  \caption{\zq{Subjective ratings on usability, usefulness, user engagement, feel in control, and user acceptance of our prototype and baseline in both scenarios. Error bars depict 95\% confidence intervals around the mean values. 
  Wilcoxon signed-rank test was employed as the data did not meet the assumptions for the parametric test.
  Significance values are reported for p < .05 (*), and p < .01 (**), abbreviated by the number of stars. We used Cohen’s d as the indicator of effect size for significant comparisons.}}
  \vspace{-2mm}
  \label{fig:overall}
\end{figure*}
\begin{figure*}[htbp]
  \centering
  \includegraphics[width=\linewidth]{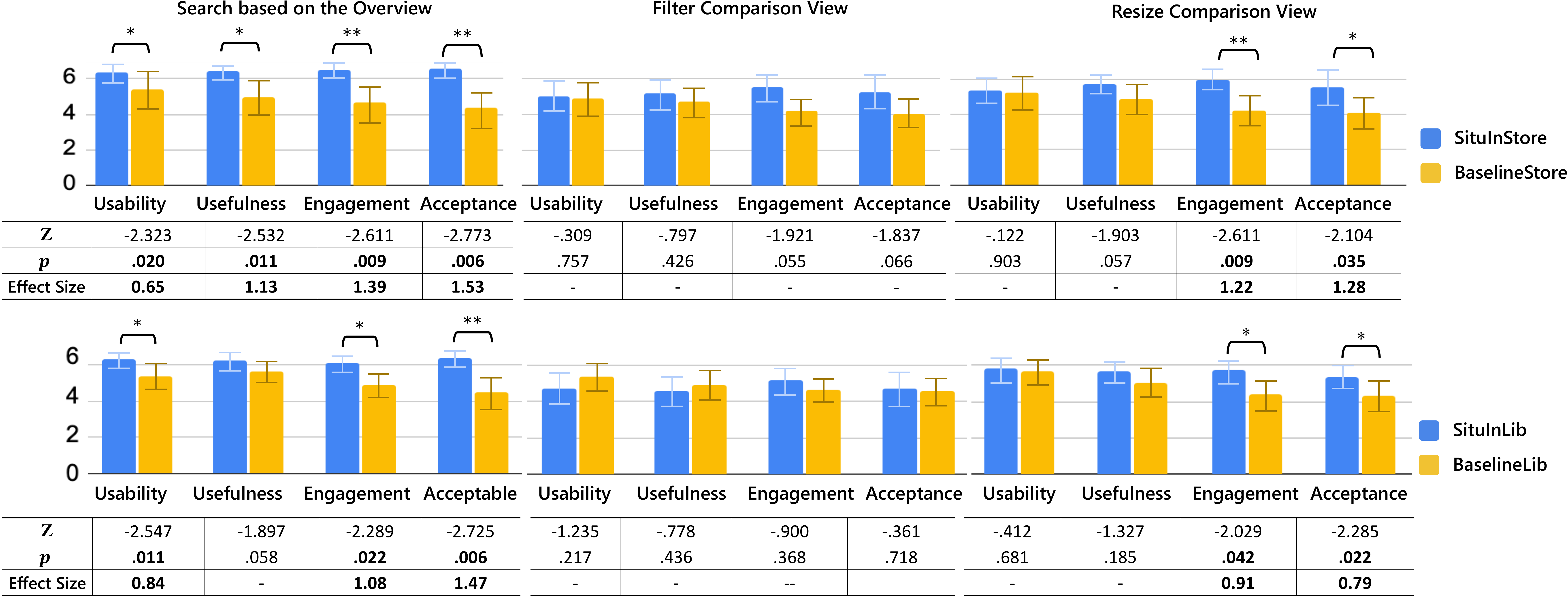}
  \vspace{-2mm}
  \caption{\zq{Subjective ratings on the interactions for different situated visualizations in both scenarios. 
  Error bars depict 95\% confidence intervals around the mean values. Wilcoxon signed-rank test was employed as the data did not meet the assumptions for the parametric test. Significance values are reported for p < .05 (*), and p < .01 (**), abbreviated by the number of stars. We calculated and presented Cohen’s d scores as indicators of effect size for significant comparisons.}}
  \label{fig:interact1}
  \vspace{-3mm}
\end{figure*}
\subsubsection{Measures and Analysis}
We gathered four types of data in the study: 
(1) participant operations in each condition with the timestamps; (2) questionnaire responses with the ratings of the interactions on usability, usefulness, user acceptance~\cite{koelle2020social}, and engagement, as well as the rankings of interaction techniques and overall ratings regarding the prototypes in each condition, using a seven-point Likert scale; (3) third-person video recordings of the participants' behaviors for reflection in the post-study interviews; and (4) audio recordings of the participants' interview responses.
For quantitative data, we used nonparametric tests, specifically, Wilcoxon signed-rank tests, to analyze the questionnaire responses as the data did not meet the assumptions for parametric tests.
Additionally, when comparing more than two types of interaction (e.g., gaze-based, hand-based, and spatially-aware object-based interactions), we conducted a Friedman test on the ratings. Post-hoc Wilcoxon signed-rank tests were used to assess specific pairwise differences between interactions.
For qualitative data from the interviews, two authors conducted thematic analysis~\cite{clarke2015thematic}. They first independently reviewed and analyzed the interview transcripts. They then engaged in two rounds of discussions to refine the results and achieve agreement. 
We triangulated qualitative findings with the quantitative results in the questionnaires when reporting our results.


\begin{figure*}[htbp]
  \centering
  \includegraphics[width=\linewidth]{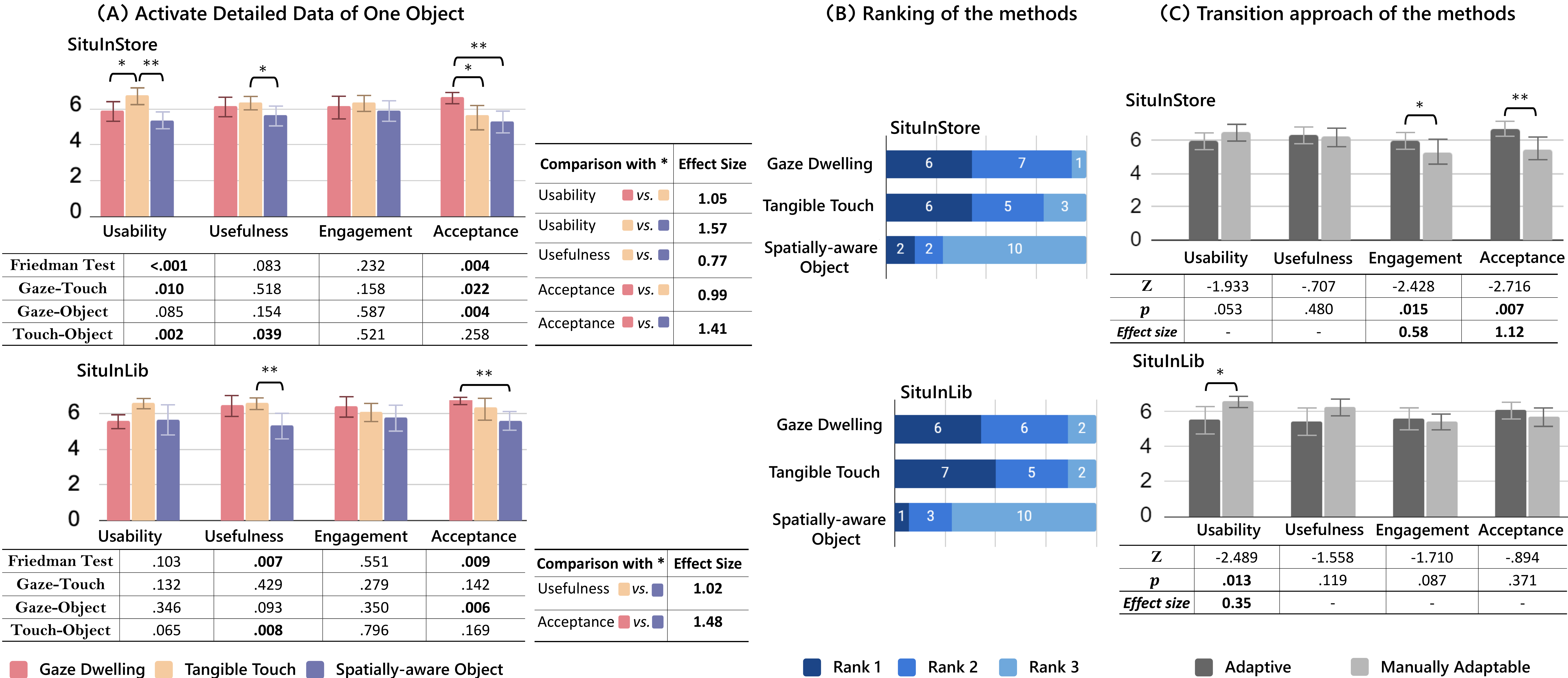}
  \vspace{-2mm}
  \caption{\zq{Subjective rating on different interactive techniques for activating detailed information of one object, the transition methods, and user rankings of these techniques. Error bars depict 95\% confidence intervals around the mean values.
  Friedman test was employed for more than two types of interactions and post hoc Wilcoxon signed-rank tests were used to assess specific pairwise differences.
  Significance values are reported for p < .05 (*), and p < .01 (**), abbreviated by the number of stars. We calculated and presented the Cohen’s d as indicators of effect size for significant comparisons.}}
  \label{fig:interact2rank}
  \vspace{-4mm}
\end{figure*}
\subsection{Quantitative Study Results}
We first present the quantitative results including the overall user perception ratings of the AR prototypes compared to the baselines, as well as detailed ratings and rankings of the proposed interactions for visualization activation, manipulation, and transition. 
Following this, we provide the number of recorded passers-by and a summary of the logged interactions in each public scenario.
Lastly, we report the qualitative feedback from participants, exploring the factors that shaped their ratings and preferences.

\subsubsection{The overall ratings of user experience.} \label{overallRating}
As depicted in Fig.~\ref{fig:overall}, participants perceived our prototype to be significantly easier to use, more useful, engaging, and acceptable compared to the baseline in the store setting. In the library scenario, they reported significantly higher scores for usefulness, engagement, and user acceptability for our prototype; however, there was no significant difference in usability scores.
The detailed statistical result is shown below the chart in Fig.~\ref{fig:overall}.

\subsubsection{User perception of different interaction designs}
Generally, participants held similar perceptions regarding the various interaction designs in both scenarios when compared to the baseline.
As depicted in Fig.~\ref{fig:interact1}, participants found that using gaze-based dwelling to search for objects based on the overview visualization (Fig.~\ref{fig:Overview} O1) was significantly easier to use, more useful, engaging, and acceptable in the public store scenario.
In the library scenario, participants reported significantly higher scores for usability, engagement, and user acceptance when interacting with the overview visualization, while no significant difference was observed in terms of usefulness.
Regarding the filtering of comparison data (Fig.~\ref{fig:Compare} C3), no significant differences were observed between our method and the baseline in both scenarios.
Furthermore, participants rated the resizing interaction for compared data (Fig.~\ref{fig:Compare} C4) with our methods significantly better than the baseline in terms of engagement and user acceptability in both scenarios.

We also gathered data on participants' perceptions and preferences for the three interactive methods used to activate detailed information, as well as the methods (adaptive and adaptable) for switching between them (Sec.~\ref{fig:Detail} E1-E3).
As illustrated in Fig.~\ref{fig:interact2rank} (A), participants found \emph{gaze dwelling} to be more acceptable for triggering detailed information in both scenarios compared to \emph{spatially-aware object} interaction. \emph{Gaze dwelling} also received higher acceptance than \emph{tangible touch} in the store.
Additionally, participants considered \emph{tangible touch} to be more useful than \emph{spatially-aware object} interaction and the library. Moreover, in the store, they reported that \emph{tangible touch} had higher usability scores compared to \emph{spatially-aware object} interaction and \emph{gaze dwelling}. 
For the edge-based slider (Sec.~\ref{fig:Detail} E4), which was kept consistent in both our prototype and the baseline for filtering the data, there was no significant difference between the two public scenarios and study conditions.

Fig.~\ref{fig:interact2rank} (B) shows that nearly half of the participants ranked \emph{gaze dwelling} or \emph{tangible touch} as their most preferred method in both scenarios. Only two participants ranked \emph{spatially-aware object} interaction as their top choice in the store and one in the library.
Regarding the two transition methods, participants provided different responses in the two public places (Fig.~\ref{fig:interact2rank} (C)). In the store, they perceived the adaptive method for switching interactions as more engaging and acceptable than the manual method (adaptable). However, in the library, the manual method was considered easier to use than the adaptive way.

\subsubsection{Number of passers-by and summary of the logged interactions} \label{QuanFootTraffic}
A total of 68 passers-by were recorded in the library, and 162 passers-by were recorded in the store during the entire study. On average, each participant encountered approximately five passers-by in the library ($MIN= 1, MAX = 12, SD = 3.01$) and around 12 passers-by in the store ($MIN = 3, MAX = 45, SD = 12.19$).
Regarding interaction logging, we observed that the number of operations in our prototypes (store: $MEAN = 27.71, SD = 2.55$, lib: $MEAN = 29.86, SD = 2.85$) was significantly higher than that in the baseline (store: $MEAN = 15.86, SD = 1.46$, lib: $MEAN = 15.57, SD = 1.70$) in both scenarios (store: $Z=-3.300, p < 0.001$, lib: $Z= - 2.731, p = 0.006$), \zq{as the eye-based interactions were triggered more frequently than others in our prototype.}


\subsection{Qualitative Feedback}
\subsubsection{\textbf{Users Preferred the Seamless and Unobtrusive Interaction in Public Environments}}
In both public scenarios, our prototype received favorable feedback from participants due to its seamless and unobtrusive interaction design. Participants described the interactions as \textit{``subtle and easily blending in with everyday movements'' (P5)} in public spaces, avoiding drawing undue attention.
Comparably, eight participants found the hand-based methods, especially mid-air gestures used in the baseline for functions like triggering the overview, to be the least acceptable. They expressed reluctance to use these gestures in similar public scenarios in the future, primarily because of the \textit{``extensive hand movements'' (P14)} involved.
Five participants particularly shared concerns that suspended hand movements would be \textit{``noticeable and make me appear strange'' (P3)}, especially when other people were nearby, potentially causing disturbances.
These findings echo the significant results in the overall experience ratings (Fig.~\ref{fig:overall}).


\subsubsection{\textbf{Environmental Impact on Interaction Experience}} \label{reEnvironmental}
Our study results suggest that users' acceptance of interactions in public is influenced not only by the design of the interaction being performed in the environment but also by various other environmental factors, such as physical constraints or uncertainty of the interactive objects.
We learned that users might hold varying perceptions of the same interaction in the two environments; even if they have similar perceptions, the underlying reasons could be different.
In the library, six participants reported that repeatedly holding and moving books, especially the thicker and heavier ones, could be physically strenuous and inconvenient. Consequently, they considered the spatially-aware object interaction to be less practical compared to the tangible touch interaction.
Similarly, participants in the store gave lower ratings to the spatially-aware object interaction. However, they attributed such scores to the larger movements required by this method. This discrepancy between the spatially-aware interaction and the tangible touch interaction is clearly evident in the significantly lower usefulness scores assigned to the spatially-aware object interaction, as illustrated in Fig.~\ref{fig:interact2rank} (A).

In contrast, participants favored hand-based tangible touch interactions in both environments, albeit for distinct reasons. 
In the library, three participants noted their inclination towards hand-based interactions, explaining that the proximity of the books (i.e., high object density) rendered eye-based interactions less convenient (\textit{e.g.,} \textit{``the books were so dense and so close to me that I did not feel the convenience of using eyes'' (P8)}).
In addition, because of the relatively low foot traffic in the library, eight participants underscored the perceived convenience and reduced reservation of hand gestures, which likely influenced their preferences. 
In the store, six participants preferred the tangible touch method because \textit{``it integrated smoothly with their typical shopping behaviors'' (P3)} and the provided objects were \textit{``easy to handle'' (P7)}.
This factor may contribute to the variability in usability scores observed in the two scenarios, as depicted in Fig~\ref{fig:overall}. Additionally, it might explain the lack of significance between the gaze dwelling and tangible touch methods in the library, in contrast to the store scene, as shown in Fig.~\ref{fig:interact2rank}.

Furthermore, the limited space between bookshelves had a notable impact on how participants perceived the adaptive transition method (Section~\ref{detailed}) for accessing detailed data of a book. 
Given the relatively close proximity between participants and the books, participants in the library tended to prefer the adaptable method, which requires manual switching of interaction modes.
The main reason is that they cannot easily adjust the distance between themselves and the bookshelf due to the limited space for spatial movement. Even though we preset the parameters of the adaptive method in the library separately, the small space caused frequent switches between different interactions, leading to an unstable interaction experience and making users confused.
In contrast, in the store, they found the adaptive approach more acceptable because it eliminated the necessity for manual mode switching, allowing for a seamless interaction experience that did not disrupt their ongoing tasks. This is evident in their ratings in Fig.~\ref{fig:interact2rank} (C).

\subsubsection{\textbf{The Convenienceof Eye-based Interaction}}
The majority of participants (10/14) preferred and tended to use gaze-based interactions in public settings. They appreciated the \textit{``strong privacy features'' (P7)} offered by this mode and demanded interactions to occur \textit{``silently'' (P2)} without attracting attention or disrupting others. Ten participants specifically valued the convenience of the eye-based interaction with the overview, including the triggering and searching for objects, because it \textit{``happens in a stealthy way'' (P1)}. This approach also helped them locate objects or understand the overall items with less effort than other methods (Fig.~\ref{fig:interact1}).
In addition, the eye-based method can be complementary to the hand-based interaction dealing with accessibility issues in real-world scenarios. Approximately half of the participants noted that in a real store, they might encounter difficulties in physically interacting with certain products like furniture or appliances. Therefore, they surmised that the gaze-dwelling interaction would likely be more accepted in such scenarios (Fig.~\ref{fig:interact2rank} (A)).

\subsubsection{\textbf{Technical Constraints Drive Preference for Simplicity and Stability}}
The post-study interviews revealed that, despite participants perceived our prototype as superior to the baseline in terms of usefulness, engagement, and acceptance in both scenarios, there were concerns about the reliability of our interactions. 

The majority of participants (9 out of 14) met the situation that the prototype failed to recognize object rotation when filtering data in Fig.~\ref{fig:Compare} C3 (Section~\ref{comparison}).
This led to repeated attempts to rotate items, which participants found unnatural and impractical. P11, for instance, mentioned, \textit{``I think it is probably because of the limited view of the AR, but it felt even more challenging in a public place. I would not want to go through this repeatedly.''}
In contrast, with the baseline, although the participants needed to raise their hands to select the filter button, which is an uncommon behavior in daily life, they perceived the operation as more brief and straightforward. 
\zq{Instances of recognition failure are infrequent, as the majority of participants (12/14) successfully filtered data after two or three attempts. However, even occasional failures can significantly impact participants' trust in the prototype and their perception and feelings as \textit{``I can not stand this happening in public.'' (P9)},} resulting in their ratings of usability, usefulness, and acceptance for the filtering of comparison view in Fig.~\ref{fig:interact1}.

The lower scores observed for spatially-aware object interactions (e.g., object rotation, translation, etc.) in terms of usefulness and acceptance in Fig.~\ref{fig:interact2rank} may be \zq{somewhat influenced by the occasional failed attempts by participants that occurred in public. However, this is not the primary reason, as participants provided insights during the interviews.}
On one hand, they displayed a clear dislike of relatively large movements in public settings, both due to social stigma and \textit{``physical efforts of moving heavy or large objects'' (P14)}. On the other hand, setting a small movement threshold for registering interaction is prone to the risk of accidental triggering, which was not well-received by the participants, \textit{``the data just suddenly popped up'' (P9)}. 
Interestingly, two participants still ranked the spatially-aware object interaction as their favorite in Fig.~\ref{fig:interact2rank} (B). They believed that it has the potential to be \textit{``a cool interactive method with gentle movements, while not drawing attention from people nearby in public'' (P10))} as the technology matures.

\subsubsection{\textbf{Familiar and Consistent Interactions for Daily Tasks}}
The tangible touch interaction was ranked as the most preferred method due to its learnability, familiarity, and quick feedback.
Three participants who had not used the AR headset specifically stressed the familiarity of this interaction compared to the gaze dwelling. They mentioned that while \textit{``the eye-gaze interaction seems more suitable for public use, honestly, I think I need some time to get used to it.'' (P6)}
Another advantage of tangible touch over gaze dwelling, in spite of the unobtrusive and hands-free nature of the latter, is its promptness as four participants commented. 
\textit{``After I get used to it, the 3-second activation delay for gaze-based interaction felt kinda long when shopping'' (P7)}. 

Another reason why some participants found \emph{tangible touch} more usable than \emph{gaze dwelling} for activating details is the desire for consistency with the subsequent analytical interactions.
Three participants mentioned that they prefer to stick with tangible touch as they need to filter the data by physically manipulating the slider attached to the object by hand (Sec.~\ref{fig:Detail}). Switching between different interaction methods can sometimes impose extra cognitive load and disrupt their tasks with situated visualization.
These points explain the significant difference between the usability ratings of the two interaction modes in Fig.~\ref{fig:interact2rank} (A).
Besides action consistency, semantic consistency is also crucial as it enables recognition rather than recall, one of the key usability guidelines~\cite{feiner1993knowledge}. Six participants reflected in interviews that resizing the comparison visualization by manipulating the distance between two objects (Fig.~\ref{fig:Compare} C4) was intuitively aligned with their perceptions and could be seamlessly integrated with the previously activated operation. As expressed by P12, \textit{``Scaling the visualization by changing the distance between objects feels very natural, intuitive, and the visual feedback that shows the distance is helpful.''}

\section{Discussion}
Our findings show the promise and limitations of our proposed interaction design with situated visualization used in the two scenarios.
In this section, \zq{we first reflect on the positioning of our work in the field of social acceptability based on the literature, and then} we discuss the design implications based on the findings, the limitations, and future work.

\subsection{Reflections on Designing and Evaluating User Acceptable Interactions with Situated Visualization}
\begin{figure}[htbp]
  \centering
  \includegraphics[width=\linewidth]{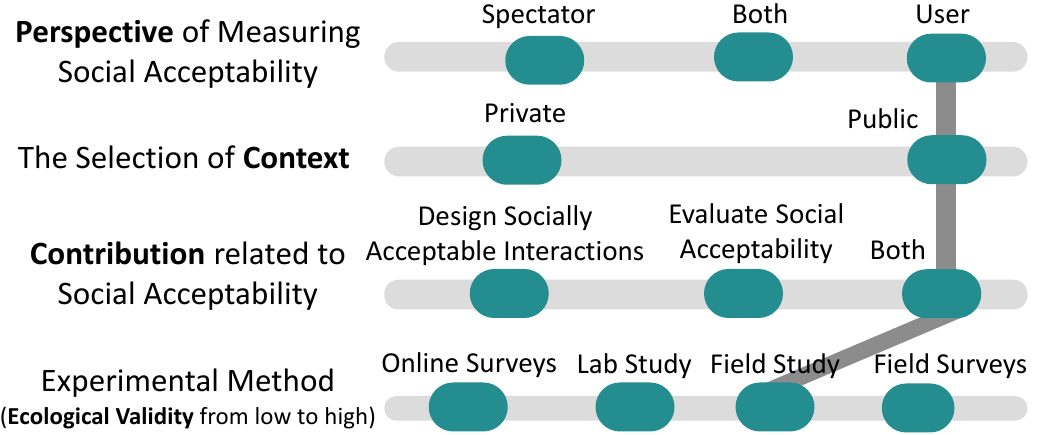}
  \caption{\zq{We reflect the aspects within the scope of social acceptability in HCI for ordinary users, represented by four key dimensions. We select and connect the considered aspects in our work using dark grey lines.}}
  \label{fig:SAResearchScope}
\end{figure}

Our study validates the necessity of considering user acceptance in the design of situated visualization interactions.
As illustrated in Fig.~\ref{fig:SAResearchScope}, we delineate the scope of social acceptability across four key dimensions~\cite{koelle2020social}, 1) perspectives of measuring the social acceptability, 2) the selection context, 3) the type of contribution, and 4) the experimental method, drawn from the existing literature. 
Our work aligns with these dimensions as we select and connect the relevant aspects. In this section, we reflect on our work by discussing these dimensions.

\subsubsection{Perspective of Measuring of Social Acceptability} ~\label{perspectiveDis}
As mentioned earlier in Sec.~\ref{relatedwork2.2.2}, there are different perspectives on measuring the social acceptability of interaction in public.
Our study concentrates on assessing user acceptance, and thus, we use the term ``user acceptable interaction'' to precisely convey the focus of our paper. We draw inspiration from Rico et al.' method by providing location descriptions and asking participants to rate their acceptability separately.
Recognizing the limitations of Rico's approach, such as not offering insights into other experiential or emotional aspects and not providing information on factors contributing to more or less social acceptability~\cite{koelle2020social}, we expanded our study questionnaire to include other aspects of users' experience (e.g., usefulness and engagement) for a comprehensive evaluation.
Additionally, we conducted post-experiment interviews to gain insights into the factors contributing to the user acceptance of interactions, and details are elaborated in Sec.~\ref{designImplication}.
It is also worth understanding and measuring the spectator's perspective of the interaction with situated visualization in public environments. We designate it as a valuable avenue for future research (Sec.~\ref{Limit&futurework}). 
We discuss several potential scenarios suitable for considering the spectator's viewpoint in Sec.~\ref{contextDis}.

\subsubsection{The Selection of Context} \label{contextDis}
Drawing from both the literature and our findings, we underscore the impact of context on users' perceptions of interaction acceptance with situated visualization.
The range of contexts could extend from private to public environments~\cite{montero2010would}. 
When using situated visualization in non-open or collaborative settings, subtle and unobtrusive interaction design principles may not be as applicable as candid interaction styles~\cite{ens2015candid}. 
For example, ensuring that the user's intention is effectively communicated to their collaborators in a collaborative data analysis project might be more important than any potential awkwardness that could arise from the interactions.
However, in face-to-face communication scenarios using situated visualization~\cite{ens2020uplift}, it can be essential to consider the potential for misunderstandings and social awkwardness arising from hand-pointing interactions directed at others.
In addition, in various public settings where users independently perform tasks like crossing the street or taking the subway~\cite{williamson2011multimodal}, people's considerations of social acceptability may prioritize safety concerns. Interaction with situated visualizations in these scenarios should be considered a secondary task, which does not occupy too much attention~\cite{lindlbauer2019context}.
Our work explores two scenarios in public where users leverage situated visualization for daily tasks independently. As people's perceptions of social acceptability may vary in different contexts, it deserves to explore other scenarios or analyze these variations in future works.

\subsubsection{Contribution related to Social Acceptability}
In terms of contributions,  prior research falls into three categories: those concerned with validating social acceptability, those emphasizing the design of socially acceptable interactions, and those considering a combination of both.
For validating social acceptability, we have discussed the literature in Sec.~\ref{perspectiveDis} based on the perspective of measurement. 
Another category of work contributes by designing socially acceptable techniques to aid users in various tasks or scenarios~\cite{lee2018designing, hsieh2016designing, lucero2014notifeye}. 
This type of work regards social acceptability as a pivotal aspect of the design process, with some studies also incorporating acceptability evaluations~\cite{hsieh2016designing, freeman2014towards, bally2012shoesense}.
Our work aligns with this category, as we followed a user-centered design approach to craft interactions deemed acceptable by users and subsequently evaluated them by collecting users' self-perceptions of acceptability in public.

Furthermore, it is crucial to highlight that designing interactions with situated visualization considering social acceptability is not a binary choice but rather a dynamic process~\cite{koelle2020social, koelle2017all}. 
As new technologies develop and gain adoption, individuals gradually become acquainted with and adapt to novel interactive behaviors when using situated visualization in reality. 
Their perceptions of social acceptability may evolve over time as they accumulate more experience and feedback in their daily lives. 
Therefore, we advocate that the design of socially acceptable interactions with situated visualization should be rooted in an understanding of users' current needs. This involves exploring nuanced demands and feedback through contextual analysis or field studies, rather than relying solely on past literature for design considerations.

\subsubsection{Experimental Method}
For experimental methods, Koelle et al. classified four types based on the literature~\cite{koelle2020social}, as shown in Fig.~\ref{fig:SAResearchScope}.
We employ a field study to ensure that our assessment of situated visualization interaction design has high ecological validity.
Nevertheless, online surveys and lab studies each have their unique advantages. 
Online surveys offer a convenient means to gather extensive user feedback, while lab studies afford greater control over the experimental process. 
Field studies, in contrast, prioritize natural settings within specific contexts to capture authentic user feedback and identify potential challenges when implementing new technologies in real-world scenarios~\cite{Kjeldskov_Paay_2012}. 
We do not insist that a field study is mandatory for designing interactions with situated visualization~\cite{kjeldskov2014worth}. The choice of experimental design should align with the specific context and tasks at hand.
Field studies require additional effort for site reservations, setup, and addressing uncertainties inherent in real-world environments.
We discuss the observed environmental influences on interactions in our experiments in Sec.~\ref{designImplication}.


\subsection{Design Implications} \label{designImplication}

\subsubsection{Considering the Impact of Uncertain Environmental Changes on Interaction Experience}
Our findings suggest that users in public settings are sensitive to their surrounding environment, which aligns with previous research~\cite{grubert2012playing, grubert2013playing}. Various factors, including the number of bystanders, the degree of environmental openness, and the available physical space, can influence how users perceive and engage in interactions. What may appear suitable based on requirements analysis or lab testing can lead to unexpected issues when deployed in real-world public environments.
Furthermore, environmental aspects such as crowd density and physical space size significantly impact how users perceive interaction techniques and their overall experience \zq{(Sec.~\ref{QuanFootTraffic})}. Therefore, designing interactions for public scenarios needs to consider user acceptability in advance and be prepared for possible changes and uncertainties within the environment.
Conducting multiple field tests can assist designers in anticipating potential environmental challenges and uncontrollable factors, enabling them to enhance the context awareness of interaction techniques. Moreover, when transitioning interactions between environments or tasks that share similarities, designers should pay attention to subtle differences, such as changes in the weight, volume, and accessibility of interaction entities. This consideration is crucial for ensuring a seamless and user-friendly experience across varying contexts.

\subsubsection{Flexibility and Complexity of Multimodal Interleaved Interactions for Everyday Data Visualization}
Our findings validate the flexibility and utility of using multimodal interactions in real-world scenarios compared to relying solely on a single modality for interacting with situated visualization. 
We demonstrate \zq{the benefit of using multimodal interaction} to meet the user need in real-world applications. In particular, we suggest leveraging multimodal interaction for accessibility concerns in real-world scenarios.
While multimodal interactions have been extensively researched in AR, our study reveals that for situated visualizations, the choice of interaction methods should be tailored to the specific type of visualization and the analytical tasks.
However, multimodal interactions can introduce complexity and increase the learning curve for users. 
To mitigate these challenges, we recommend implementing intelligent, context-aware transitions between interaction modalities that seamlessly accommodate users' current tasks and facilitate fluid interaction and analysis of situated visualizations~\cite{zhang2020investigating}. Moreover, when implementing transition methods, maintaining consistency from activating the visualization to performing analytical operations is crucial for a cohesive user experience.
Combining adaptive with adaptable could lead to higher usability in response to complex and uncertain environments.

\subsubsection{Tailoring Interaction to User Acceptance and Environmental Constraints}
Our findings indicate that users can be reluctant to modify their regular actions in public environments to interact with situated visualizations, even when presented with familiar and stable interactions (e.g., pinch or mid-air gestures)~\cite{walter2013strikeapose}, as such actions can be perceived as awkward and unconventional in public settings.
Therefore, we recommend that the design of interactions for real-world applications should not solely rely on cutting-edge recognition technology but also consider users' willingness and the constraints within the environment. 
However, it is crucial to acknowledge \zq{interactions should be usable first and then deemed acceptable by users.} Even discreet and subtle interactions, if not easy to manipulate, can cause user discomfort in public due to repeated attempts to use them.

\subsubsection{Drawing on Real-Life Experience and Experiments to Design Interaction for Situated Visualzation}
Lave's work in \textit{Cognition in Practice} elucidates that conducting user research in a laboratory setting does not necessarily provide more meaningful guidance than having individuals describe their lives~\cite{lave1988cognition}. 
The work underscored the value of studying people's needs and behaviors in the field, describing it as ``cognition in the wild''.
It is necessary to draw on users' real-life experiences before designing the interaction for situated visualization, which is a visualization form that is closely tied to real-world environments.
Additionally, it is crucial to consider how to integrate users' natural behaviors and context into the interaction. These require careful and in-depth observation and exploration of users' behaviors before developing and designing interactions.
Furthermore, we suggest involving target users earlier in the design process through participatory design or co-creation, from the beginning and not as an afterthought in evaluation.
This can avoid investing expensive implementation efforts in the interaction that users might never accept.

\subsubsection{Benefits and Risks of Eye-based Interactions for Situated Visualiztion}
In previous studies on situated visualization, eye-based input has been employed to measure user attention or evaluate embedding visualization in real-world context~\cite{kurzhals2022evaluating, chen2023iball}. 
We find that the eye-based interaction was more acceptable because of the implicit and less obvious appearance in public.
In addition, eye-based interaction is suitable for combining with other interactions for situated analysis, to express people's implicit intentions or activate analytical operations. Such combinations should be designed carefully to leverage the complementary effect of eye-based input with other input modalities.
However, our findings also disclose several drawbacks of eye-based interaction, such as instability and lack of physical feedback of eye-based interaction. Therefore, designing eye-based input as an implicit form of interaction necessitates a careful distinction between intentional and unconscious eye inputs to avoid the \textit{Midas Touch} problem~\cite{xiao2018mrtouch, jacob1990you}. 
In addition, intentional eye-gaze input should not be overly frequent, as it can lead to user fatigue. Lastly, when employing eye-based interactions in public settings, it is essential to consider privacy issues and the inadvertent exposure of user interests.


\subsection{Limitations and Future Work} \label{Limit&futurework}
There are several limitations in our work. 
First, when designing and measuring interactions, we only took into account the user's perception and feedback, without considering the perspective of bystanders in the design and validation of the interactions~\cite{reeves2005designing}.
Future studies can explore the social acceptability of interactions for situated visualization more comprehensively by taking the spectators' view.
\zq{}
Second, to discover the limitations and benefits of the proposed interaction, we evaluated the interaction techniques through concrete routine tasks rather than the exploratory study. Future work requires exploring how people would like to use the different interaction techniques without giving clear goals or tasks.
Third, we found that designed interactions may be perceived as more socially acceptable than they are because the AR glasses were inevitably shown to the people around the participant, which somehow explains the user's interaction intent to those around them, which can be seen as a way to improve the social acceptability~\cite{ens2015candid, koelle2020social}. 
In the future, when AR glasses tend to be more discreet, users might have higher expectations for interactions when leveraging the situated visualization for personal use in public settings. 
Fourthly, due to limitations imposed by current devices, some user-generated ideas for interactions were unable to be implemented in the development process. For instance, one participant suggested an implicit pinch method as an alternative to the existing mid-air gestures that require users to lift their hands, which might be simpler and more efficient than the current design.
Therefore, with the development of immersive technologies and the iteration of devices, it may be necessary to evaluate new devices or more stable technologies in the future.
\section{Conclusion}
This study explored the design of interaction techniques with situated visualization for daily tasks in public environments.
Through a formative study, we concretize users' contextual needs of situated visualization in two common public spaces, including the data, display, representation, and interaction. 
Then, we summarized the potential interaction input modalities that are publicly acceptable and technically feasible in public, in conjunction with the required visualization. 
We organized an iterative design process to progressively explore, implement, and improve the interaction techniques for situated visualization. 
Our evaluation results suggest that the proposed interaction is publicly acceptable, flexible, and practical in real-world public scenarios. While highlighting the potential of multimodal interaction in public settings, we emphasize the importance of considering factors such as user acceptance, interaction complexity, and environmental dynamics.
We believe that considering the context of using interactions prior to design and implementation is critical to facilitating more practical and acceptable human-data interactions in real-world environments.

\begin{acks}
This work is supported by the Research Grants Council of the Hong Kong Special Administrative Region under General Research Fund (GRF) with Grant No. 16207923.
\end{acks}
\bibliographystyle{ACM-Reference-Format}
\bibliography{sample-base}










\end{document}